\documentclass[12pt]{article}
\usepackage{latexsym, amsmath, epsfig}
\DeclareFontFamily{OT1}{rsfs10}{}
\DeclareFontShape{OT1}{rsfs10}{m}{n}{ <-> rsfs10 }{}
\DeclareMathAlphabet{\mathscript}{OT1}{rsfs10}{m}{n}

\if@twoside\oddsidemargin25mm
\else\oddsidemargin25mm\evensidemargin25mm\marginparwidth25mm\fi
\def\hat{\widehat}

\def\bpl{\Big(}
\def\bpr{\Big)}

\def\t{\theta}

\def\w{\omega}

\def\der{\partial}
\def\brr{\begin{equation}}
\def\err{\end{equation}}
\def\brr{\begin{eqnarray}}
\def\err{\end{eqnarray}}
\def\ba{\left(\begin{array}}
\def\ea{\end{array}\right)}

\def\Cslash{\hbox{\ooalign{$\displaystyle{C}$\cr\hspace{.3mm}$/$}}}

\newcommand{\dr}{\raise.3ex\hbox{$\stackrel{\leftarrow}{\partial }$}{}}
\newcommand{\dl}{\raise.3ex\hbox{$\stackrel{\rightarrow}{\partial}$}{}}

\newcommand{\ft}[2]{{\textstyle\frac{#1}{#2}}}
\newcommand{\ns}{\normalsize}
\renewcommand{\theequation}{\thesection.\arabic{equation}}

\renewcommand{\a}{\alpha}
\renewcommand{\b}{\beta}

\renewcommand{\d}{\delta}

\begin{document}


\begin{titlepage}

\vspace{-3cm}

\title{
   \hfill{\ns HUB-EP 99/10\\} 
   \hfill{\ns UPR-835T\\} 
   \hfill{\ns hep-th/9903028\\[1cm]}
   {\LARGE Intersecting Orbifold Planes and Local
    Anomaly Cancellation in {\it M}-Theory}\\[.5cm]}
                       
\author{{\bf
   Michael Faux$^{1}$, \,
   Dieter L{\"u}st$^{1}$ \,
   and Burt A.~Ovrut$^{2}$}\\[5mm]
   {\it $^1$Institut f\"ur Physik, Humboldt Universit\"at} \\
   {\it Invalidenstra\ss{}e 110, 10115 Berlin, Germany} \\[3mm]
   {\it $^2$Department of Physics, University of Pennsylvania} \\
   {\it Philadelphia, PA 19104--6396, USA}}
\date{}

\maketitle

\begin{abstract}
\noindent
A systematic program is developed for analyzing and cancelling local anomalies 
on networks of intersecting orbifold planes in the context of
${\it M}$-theory.  Through a delicate balance 
of factors, it is discovered that local anomaly matching on the
lower-dimensional intersection of two orbifold planes may
require twisted matter on those planes which do not
conventionally support an anomaly (such as odd-dimensional planes).
In this way, gravitational anomalies can, in principle,
tell us about (twisted) gauge groups on subspaces which are not necessarily
ten-, six- or two-dimensional.  An example is worked out for the case
of an $S^1/{\bf Z}_2\times T^4/{\bf Z}_2$ orbifold and possible 
implications for four-dimensional physics are speculated on.
\end{abstract}

\thispagestyle{empty}

\end{titlepage}


\section{Introduction}
Anomalies have been curiously adept at providing insight into fundamental
concepts and indicating new phenomena.  The role \cite{greenschwarz}
of local gauge
and gravitational anomaly cancellation \cite{agwitten}
in fomenting the so-called
first superstring revolution is well known. But the ongoing development
of nonperturbative ``string" dynamics has also relied strongly on 
constraints imposed by anomaly cancellation in effective field theories.
Since the nonperturbative picture is now understood to involve 
an eleven-dimensional description, eleven-dimensional supergravity
has become a central tool in the exploration
of the yet-mysterious underpinnings of whatever 
dynamics comprise {\it M}-theory.  Anomalies have an important
place in this story.

In this paper, we develop a systematic method for analyzing {\it M}-theory 
orbifold anomalies in situations where there are various orbifold 
planes of different dimensionalities which intersect.  
Particularly, we indicate a way in which the requirement of 
{\it local} anomaly cancellation
on subspaces corresponding to the intersection of two orbifold
planes can require twisted matter propagating on the entirety
of one of the planes regardless of whether that
plane actually supports a separate local anomaly.  This works
because fermions propagating on an intersecting plane can couple 
to currents localized on the intersection via projections which are 
chiral from the point of view of the (lower-dimensional) intersection.
The associated contributions to the anomaly are distinguishable
by virtue of divisors which properly modify standard index theory
results in these situations.

Eleven-dimensional supergravity has a solitonic superfivebrane with
chiral zero-modes.  Hence, fivebrane worldvolume gravitational anomalies 
posed an early puzzle to the consistent realization of {\it M}-theory
effective actions. 
As anticipated in \cite{vw} and realized in \cite{dlm}, the cancellation
of these anomalies requires an extension of the minimally coupled
supergravity action in the form of a coupling $GX_7$, where 
$G$ is the four-form field strength and $X_7$ is a seven-form involving the
eleven-dimensional Riemann tensor.  The chiral modes on the fivebrane 
worldvolume also couple to the normal bundle involving $SO(5)$ 
spacetime symmetries not broken by the presence of the fivebrane.
The cancellation of associated $SO(5)$ anomalies posed a further
puzzle which was analyzed and partially solved in \cite{wfb} and 
more recently resolved in \cite{fhmm} by a subtle mechanism
involving the $CGG$ interaction present in the 
minimally coupled theory.

Chiral anomalies also arise in orbifold compactifications of 
{\it M}-theory.  Such constructions provided some of the initial 
impetus for the current faith in a unified description of the five 
erstwhile separate ten-dimensional string theories, and thereby have 
played a role in the so-called second superstring revolution.
Generally, orbifolding eleven-dimensional supergravity 
removes from the bulk gravitino and the bulk three-form potential all but
a chiral projection on any even-dimensional hyperplane fixed
by the action of the discrete group that defines the orbifold.
In cases where there are ten- or six-dimensional fixed planes,
the cancellation of gravitational anomalies induced by couplings involving
these projections poses yet more puzzles to
the consistent realization of {\it M}-theory orbifolds, only some of which
have been completely resolved.

The prototype {\it M}-theory orbifold was described 
by Ho{\v r}ava and Witten in \cite{hw1, hw2},
where the connection between eleven-dimensional supergravity and
the $E_8\times E_8$ heterotic string was first indicated.  
In this case, one of the spatial dimensions was compactified
on $S^1/{\bf Z}_2$, which is an orbifold with two ten-dimensional 
fixed hyperplanes.  In this case cancellation of induced gravitational 
anomalies necessitates the presence of ten-dimensional $E_8$ gauge matter 
propagating on each of the two ten-planes.  These, of course, provide
the two $E_8$ factors known previously from the perturbative 
heterotic string perspective, in which the eleventh dimension is
re-interpreted as the (small) string coupling constant.

A central aspect of the Ho{\v r}ava-Witten analysis was the requirement
of anomaly cancellation independently at each point in 
eleven-dimensional spacetime.  Since the one-loop anomalies in that case
derive from from the coupling of the {\it eleven}-dimensional gravitino to 
{\it ten}-dimensional currents, the anomaly is characterized by
expressions which differ from those obtained from usual index theorems.
This difference is a specific factor of 1/2 derived from the
fact that there are {\it two} orbifold planes. 

One would like to have a guiding principle for constructing realistic models
of nature from {\it M}-theory.  In perturbative string theory, the
requirement of modular invariance was eventually
understood to imply anomaly freedom.
In {\it M}-theory we have no such principle which we can point to 
which offers an ``explanation" for anomaly freedom.  But we do
have the requirement of anomaly freedom itself.  This turns out
to be a powerful tool in its own right.
It is conceivable that anomaly cancellation in effective theories are
somehow {\it equivalent} to the microscopic consistency 
requirements related to a fundamental description of {\it M}-theory.

In this paper, we offer a concise and self-contained account of the
anomaly cancellation issues associated with three ``basic"
{\it M}-theory constructions:
the {\it M}-fivebrane, the $S^1/{\bf Z}_2$ orbifold, 
and the $T^5/{\bf Z}_2$ orbifold.  These are reviewed using 
the specific tools used afterward to analyze the more
interesting $S^1/{\bf Z}_2\times T^4/{\bf Z}_2 $ orbifold,
and are presented in detail for the reason that
the same calculations involved constitute necessary sub-analyses
in the latter case. The {\it M}-fivebrane in particular 
is central to all phenomenological
constructions based on {\it M}-theory. 
Furthermore, the independent presentations of these situations 
using our specific tools enable us to focus on individual
generic aspects such as the concept of anomaly inflow and
the issue of ``wandering fivebranes"  which can mediate phase transitions.

In section 2 we review the cancellation of the worldvolume gravitational
anomaly for the {\it M}-fivebrane.  This allows us to motivate
and introduce the basic tools common to anomaly analyses
for all {\it M}-theory orbifolds, and also to set our notation.
The basic tools are the anomaly polynomials mentioned above, the
$CGG$ and $GX_7$ Chern-Simon's interactions and special objects
called brane-currents important for describing magnetic and 
electric sources for $G$ concentrated on sub-manifolds.  

In section 3 we review the cancellation of gravitational and
gauge anomalies in the context of the Ho{\v r}ava-Witten $S^1/{\bf Z}_2$
orbifold.  This entire analysis carries over as a sub-analysis
in the more complicated case analyzed in section 6.  Numerical 
details derived in section 3 are essential to the analysis of
section 6.

In section 4 we review the cancellation of the gravitational anomaly
in the context of the $T^5/{\bf Z}_2$ orbifold \cite{dasmuk, wittens5}.  
This scenario comprises another sub-analysis for 
the case studied in section 6, but an orthogonal one.   
The presentation in this section also allows useful parallels 
to be drawn in section 6 relating to paradoxes which are discussed
in that section. 

In section 5 we explain an effect not present in the case of the
${\bf Z}_2$ orbifold anomalies, but which is essential to 
anomalies in orbifolds with intersecting fixed-planes.
This ``I-brane" effect mirrors a synonymous effect in the 
context of intersecting D-branes, and involves an interplay
between electric and magnetic sources for $G$ supported
on separate but intersecting orbifold planes.  The effect
forms a crucial ingredient to the analysis of section 6.

Section 6 involves a detailed local anomaly analysis in the case
of a $S^1/{\bf Z}_2\times T^4/{\bf Z}_2$ orbifold.
This orbifold is essentially an amalgamation of the 
two ${\bf Z}_2$ orbifolds reviewed in sections 3 and 4, but there are
important new features.  The first one is that, due to the larger
discrete group, the bulk supersymmetry is halved again
on the fixed-plane intersections.  This allows for 
a richer twisted sector, enables a local Green-Schwarz
mechanism precluded in the former cases and involves the interesting
I-brane effect described in section 5.  Most importantly, the
analysis indicates how one can infer seven-dimensional 
twisted states based on their relationship to anomalies on
the six-dimensional subspaces corresponding to the intersections
with the ten-planes. There is a paradox, however, which 
leaves an unanswered question.  
As far as we are aware, ours is the first analysis into the {\it local}
aspects of anomaly cancellation in {\it M}-theory orbifolds
other than the two ${\bf Z}_2$ orbifolds described above.

In section 7 we conclude and speculate on implications our results
may have for even more complicated orbifolds related to 
realistic four-dimensional physics.  

We also include three appendices.  

Appendix A includes tables describing some results from group 
theory required by the analysis described in section 6.    

Appendices B and C serve as brief encyclopedias
to the relevant anomaly polynomials used in the main part
of the paper.  These appendices are included for reference
purposes and are meant to be practical and concise.  For this 
reason the relationship of these polynomials to
index theorems is de-emphasised.
Appendix B includes the polynomials relevant to
ten-dimensional anomalies.  This also includes two well-known ``classic"
examples from ten-dimensional supergravity to illustrate how these
same polynomials are used in more straightforward analyses.  
These are included for the benefit of readers without much
familiarity with anomaly polynomials.  Appendix C includes
a similar encyclopedia involving the analogous six-dimensional
anomaly polynomials and also includes examples.

\setcounter{equation}{0}
\section{The {\it M}-Fivebrane and Brane-Currents}
The {\it M}-fivebrane describes a prototypical spacetime defect 
supporting a potential anomaly.  This six-dimensional object has
worldvolume fields transforming as a D=6 N=2 tensor multiplet
\footnote{In our conventions, $N=2$ describes 16 supercharges, and is
therefore {\it twice} the minimum in six-dimensions.  The same superalgebra
is sometimes called $N=4$ in the literature because it corresponds to
$N=4$ in four dimensions.}.  This 
includes five scalars parameterizing the position of the fivebrane 
in eleven-dimensional spacetime as well as an anti self-dual two-form 
(ie: the three-form field strength satisfies $H\!=\!-\!*\!H$)
and a pair of antichiral spin 1/2 fields.  The tensor and the fermions
each contribute to a gravitational anomaly at one loop.  
This anomaly is characterized
by an eight-form which can be computed via the relation
\brr I_8({\rm 1\,loop})=-I_{\rm GRAV}^{\,({\rm 3-form})}(R)
     -2\,I_{\rm GRAV}^{\,(1/2)}(R) \,,
\label{ada}\err
where the minus signs indicate anti self-duality and negative chirality
respectively, and the absolute values of the two coefficients
reflect the multiplicities of the indicated fields.  The polynomials
$I_{\rm GRAV}^{({\rm 3-form})}(R)$ and $I_{\rm GRAV}^{(1/2)}(R)$ encode
the contributions to a six-dimensional gravitational anomaly due to a 
single self-dual tensor field and a single chiral spin 1/2 fermion.
These are determined by index theorems and given explicitly
in appendix C as equation (\ref{grav6}).
Substituting the polynomials given in (\ref{grav6}) into the 
expression (\ref{ada}) we easily determine 
\brr I_8({\rm 1\,loop})=\frac{1}{(2\pi)^3\,4!}\,\bpl\,
    \ft18{\rm tr}\,R^4-\ft{1}{32}({\rm tr}\,R^2)^2\,\bpr \,,
\label{fbloop}\err
where $R$ is the six-dimensional Riemann tensor expressed
as an $SO(5,1)$-valued two-form.  

The worldvolume fermions also couple to $SO(5)$ currents associated with the 
normal bundle.  This gauge group is inherited from the 
``bulk" $SO(10,1)$ diffeomorphism group, which is broken by the presence
of the fivebrane to $SO(5,1)\times SO(5)$.  Whereas the $SO(5,1)$ 
transformations include the fivebrane worldvolume diffeomorphisms,
with one-loop anomaly described by (\ref{fbloop}), the $SO(5)$
anomaly is more subtle.  Its cancellation was resolved in \cite{fhmm},
and involves mathematics which we will not need or describe in this paper.
The normal bundle anomaly can be considered independently.  We
therefore suppress this issue in the balance of this paper.

The one-loop anomaly (\ref{fbloop}) is canceled via ``inflow" from
classical variation of the eleven-dimensional action.  The classical
variation can include an anomalous contribution localized on the 
fivebrane worldvolume provided the four-form $G$ of eleven-dimensional
supergravity couples magnetically to the fivebrane
via modifications to the $dG$ Bianchi identity.
The anomaly inflow arises specifically 
due to the variation of the following terms in the action
\footnote{The normalizations are chosen as follows.  
We start with the conventions employed in \cite{fhmm}, and apply a
further scaling $C\rightarrow 2\pi\,C$ (so the object we call $C$ is
the one called $\Cslash_3$ in \cite{fhmm}).
This determines the coefficient of the 
first term to be $-\pi/3$.  With this scaling, quanta of $G$-flux
are measured in units of $\int G$ rather than $(2\pi)^{-1}\int G$,
which also explains why there is no factor of $2\pi$ in equation 
(\ref{fbbianchi}).},   
\brr S=\cdots-\frac{\pi}{3}\,\int C\wedge G\wedge G
    +\int G\wedge X_7 \,.
\label{action}\err
The $CGG$ interaction is required by the minimally-coupled 
supergravity while the $GX_7$ term is an additional higher-derivative 
interaction required by the fivebrane anomaly cancellation.
This also requires that $X_7$ transform into a total derivative under
local $SO(10,1)$ Lorentz transformations as $\d X_7=dX_6^1$.  
The precise form of $X_7$ is dictated by the anomaly cancellation.

\vspace{.1in}
\noindent
{\it A magnetic source for G:}\\[.1in]
In the presence of a single fivebrane, the four-form $G$ 
satisfies the Bianchi identity
\brr dG=\d^{\,(5)}_{W^6} \,,
\label{fbbianchi}\err
where the five-form $\d^{\,(5)}_{W^6}$ is, in the terminology of
\cite{cy}, a brane-current with support localized on
the fivebrane worldvolume, $W^6$.  Such objects have 
received critical attention in various papers, notably
\cite{wfb, fhmm, cy}, and have an interesting and somewhat involved
mathematical description.  The essential features 
necessary for our purposes are summarized as follows.

An ($\,11\!-\!d\,$)\,-\,form $\d_{M^d}^{(11-d)}$ is a brane-current if it has
localized support on the $d$-dimensional defect $M^d$ and if it 
satisfies properties of being closed, 
integrating to one over any normal fiber of the embedding space, and 
performing as a generalized Dirac delta function
by collapsing integrals to the support manifold,
$\int\d_{M^d}\wedge \Phi=\int_{M^d}\Phi$.
Furthermore, provided that manifolds $M^{d1}$ and $M^{d2}$ intersect
transversally, the
product of two associated brane-currents is itself a brane-current.  
Thus, if $M^{d1}$ and $M^{d2}$ intersect at right angles in eleven-dimensions,
\brr \d_{M^{d1}}^{(11-d1)} \wedge \d_{M^{d2}}^{(11-d2)} \equiv 
     \d_{M^{d1}\cap M^{d2}}^{(11-{\cal I})} \,,
\label{prop}\err
where ${\cal I}=d1+d2-11$ 
is the dimensionality of the transversal intersection
$M^{\cal I}=M^{d1}\cap M^{d2}$.
Subtleties involving non-transversal intersections are discussed and
resolved in \cite{cy}.  

A fivebrane coupled as in (\ref{fbbianchi}) will contribute one
unit of $G$-flux.  This is seen by integrating (\ref{fbbianchi}) over
the five dimensions transverse to the fivebrane worldvolume
using a region bounded by a four-cycle $\w$ which encompasses the 
fivebrane. 
Stokes theorem then allows us to express
the left-hand side of the integrated Bianchi identity as an integral
over $\w$, while the right-hand side is unity by virtue of a defining
property of the brane-current $\d^{(5)}_{W^6}$.  Thus, $\int_\w G=1$.

\vspace{.1in}
\noindent
{\it Anomaly inflow:}\\[.1in]
It is straightforward to determine the variation of the $GX_7$ term.
Using an integration by parts and applying the Bianchi identity
(\ref{fbbianchi}) we determine 
\brr \d\bpl\int G\wedge X_7\,\bpr=-\int_{W^6} X_6^1 \,.
\label{x7fb}\err
Note that the brane-current $\d^{(5)}_{W^6}$ included in 
the Bianchi identity collapses the
eleven-dimensional integral to a six-dimensional integral over $W^6$.
The $GX_7$ inflow contribution is characterized by the closed
gauge-invariant eight-form $X_8$ which gives rise to 
(\ref{x7fb}) upon descent. Thus we can write
\brr I_8(GX_7)=-X_8 \,,
\label{gx7}\err
where $X_8\equiv d X_7$ and $\d X_7=d X_6^1$.  This inflow contribution
must cancel against the one-loop anomaly given in
(\ref{fbloop}).

\vspace{.1in}
\noindent
{\it The total anomaly:}\\[.1in]
The total worldvolume gravitational anomaly is given by the sum of the
one-loop anomaly (\ref{fbloop}) and the inflow contribution
(\ref{gx7}). Thus, $I_8({\rm total})=I_8({\rm 1\,loop})+I_8(GX_7)$.
We require that this total anomaly vanish.  Using equation 
(\ref{gx7}), this indicates that $X_8$ must be equal to $I_8({\rm 1\,loop})$,
which is given as (\ref{fbloop}).  Thus,
\brr X_8=\frac{1}{(2\pi)^3\,4!}\,\bpl\,
    \ft18{\rm tr}\,R^4-\ft{1}{32}({\rm tr}\,R^2)^2\,\bpr\,.
\label{x8def}\err

Note that the $CGG$ term does not contribute to the fivebrane 
gravitational anomaly.  It does, however, contribute importantly
to the normal bundle anomaly (which we are suppressing).  This is
described in \cite{fhmm}.  In more general contexts, such as the 
orbifolds analyzed below, the $CGG$ term 
provides a crucial ingredient and cannot be neglected.

\vspace{.1in}
\noindent
{\it Comments:}\\[.1in]
The above analysis is easily generalized to the presence of 
any number of unit-charge fivebranes.  
If there are $N_5$ fivebranes, then (\ref{fbbianchi}) is
replaced with
\brr dG=\sum_{i=1}^{N_5}\d^{\,(5)}_{W^6_i}
\label{sumbianchi}\err
where $W^6_i$ is the worldvolume of the $i$th fivebrane.
In this case,
each fivebrane will support an independent set of worldvolume
fields (comprising a D=6 N=2 tensor multiplet)
which contributes to a one-loop anomaly localized on $W^6_i$.
At the same time there will be an inflow contribution
to the total anomaly concentrated on each fivebrane due to the sum of
terms in (\ref{sumbianchi}), and arising due to the variation of the
same $GX_7$ term derived above.  The one-loop anomaly and the inflow
anomaly will cancel each other independently on each fivebrane.  

Since eleven-dimensional supergravity has unit-charge
fivebranes as a solitons, then, by virtue of the above discussion,
any quantum theory which has eleven-dimensional supergravity
as its low-energy description should involve as well the
particular $GX_7$ interaction.
This term also proves crucial to the unraveling 
of further puzzles presented upon orbifold compactification.

\section{{\it M}-theory on ${\bf R}^{10}\times S^1/{\bf Z}_2$}
The simplest orbifold of eleven-dimensional supergravity
has had a significant impact. This construction 
\cite{hw1, hw2} provided simple answers to longstanding 
puzzles associated with the role of eleven-dimensions in the scheme 
of string-theory, provided a satisfying picture of the 
strongly-coupled dynamics of the $E_8\times E_8$ heterotic string
and has opened the door to 
a wealth of new ideas pertaining to nonperturbative fundamentals
in physics.  One thing that makes this construction so powerful
is the ease with which some of the essential calculations can
be done, a common feature of orbifold models.   

Orbifolds are defined by discrete projections which act both
on the spacetime manifold and on the field content of the theory.
These define a set of invariant hypersurfaces, known as
orbifold planes, which provide the main focus for analytical attention.
In the case of the $S^1/{\bf Z}_2$ orbifold ,
there are two {\it parallel} ten-dimensional orbifold planes within
eleven dimensional spacetime. 
Our primary interest in this paper is in more complicated orbifolds 
which involve intersecting networks of fixed-planes.  
Nevertheless, much of the analysis in those cases is identical to
that associated with simpler orbifolds without fixed-plane
intersections. So in this section we briefly but completely review 
the anomaly analysis for the case of the $S^1/{\bf Z}_2$
orbifold.  Since this analysis uses the same tools which
we will need later on, this enables a necessary appreciation 
for the some of the computational mechanics used in more general
situations.  It also supplies results
which carry over to other orbifolds whose orbifold group
includes a similar projection as a subgroup.

\vspace{.1in}
\noindent
{\it The structure of the orbifold:}\\[.1in]
Start with eleven-dimensional supergravity on 
${\bf R}^{10}\times S^1/{\bf Z}_2$.  The ${\bf R}^{10}$ factor is
parameterized by $x^A\equiv\{x^1,...,x^{10}\}$, while the one compact dimension
is parameterized by $x^{11}$, which takes values on the interval
$[-\pi,\pi]$ with endpoints identified.  Truncate the theory via
a ${\bf Z}_2$ projection which acts on the compact dimension as 
$x^{11}\rightarrow -x^{11}$.  There are then {\it two} ten-dimensional
hyperplanes fixed by this projection, namely the surfaces defined
by $x^{11}=0$ and $x^{11}=\pi$.

Demanding ${\bf Z}_2$-invariance of
the eleven-dimensional interaction term $CGG$
implies that the three-form $C$ is odd under
the ${\bf Z}_2$ projection described above. 
Thus, $C_{ABC}\rightarrow -C_{ABC}$, 
where $A,B,C\in \{1,...,10\}$.
The components $C_{ABC}$ therefore vanish on the 
fixed hyperplanes.  
However, $C_{(11)AB}\rightarrow C_{(11)AB}$.  
Therefore, from the point of view
of the two ten-dimensional fixed-planes, the three-form $C$
contributes one nonvanishing tensor $C_{(11)AB}$.  

Half the supersymmetries of the eleven-dimensional theory are broken
on the two ten-planes by the ${\bf Z}_2$ projection. 
Thus, on the fixed-planes
the one tensor $C_{(11)AB}$ organizes
along with the other fields surviving the ${\bf Z}_2$ projection
into a D=10 N=1 supergravity multiplet. This constitutes the 
``untwisted sector" of the orbifold.

We allow for a ``twisted sector" involving ten-dimensional Yang-Mills
multiplets in the adjoint representation of some gauge group
${\cal G}_i$ propagating on the $i$th ten-plane.
As reviewed below, anomaly cancellation uniquely selects this group.

\vspace{.1in}
\noindent
{\it The one-loop anomaly:}\\[.1in]
A gravitational anomaly arises due to the coupling 
of chiral projections of the bulk gravitino to currents
localized on the two fixed-planes.  
Since the two ten-planes are indistinguisable
aside from their position, this anomaly is similar on each 
of the two, and can be computed using standard formulae if
proper care is used.  The reason why extra care is needed is that
the anomaly in question actually derives from the coupling of 
{\it eleven}-dimensional fermions to {\it ten}-dimensional currents,
whereby standard index theorem results (such as those described in
appendix B) can be applied directly only in the small-radius
limit when the two fixed-planes coincide.  This is because it
is only in this limit that we describe
ten-dimensional fermions coupled to ten-dimensional currents.  
Thus, the gravitino-induced anomaly on a given ten-plane is
{\it one-half} of that described by the index theorem results 
using the (ten-dimensional) untwisted spectrum described above.

The gaugino fields living in ``twisted" Yang-Mills multiplets
also contribute a gravitational anomaly, as well as mixed
and a pure-gauge anomalies.  However, since the twisted fields are
ten-dimensional these can be computed directly using the standard formulae
(without multiplying by 1/2).

It is then straightforward to compute the one-loop anomaly using
the formulae described in appendix B.  
The local anomaly on the $i$th fixed-plane is characterized 
by the following twelve-form,
\brr I_{12}({\rm 1\,loop})_i &=& \ft14\,\bpl\,
     I_{{\rm GRAV}}^{(3/2)}(R)
    -I_{{\rm GRAV}}^{(1/2)}(R)\,\bpr
    \nonumber\\[.1in]
    & & 
    +\ft12\,\bpl\,n_i\,
    I_{{\rm GRAV}}^{(1/2)}(R)
    +I_{{\rm MIXED}}^{(1/2)}(R,F_i)
    +I_{{\rm GAUGE}}^{(1/2)}(F_i)\,\bpr \,,
\err
where the various constituent polynomials are
given explicitly in (\ref{grav10}) and (\ref{gauge10}), and
$n_i$ is the dimension of the adjoint representation of 
${\cal G}_i$.  As explained in appendix B, each term includes a 
factor of 1/2 because the relevant fermions are Majorana-Weyl
(having {\it half} the degrees of freedom of a Weyl spinor)
while the first two terms obtain an additional factor of 1/2 
(accounting for an overall coefficient of 1/4) 
for the reasons described above.
Using (\ref{grav10}) and (\ref{gauge10}) we easily compute the 
polynomial $I_{12}({\rm 1\,loop})_i$.  The result has a 
${\rm tr}\,R^6$ term, with coefficient proportional to 
$(n_i\!-\!248)$.  However, the anomaly can only be cancelled if the
twelve-form $I_{12}({\rm 1\,loop})_i$ factorizes as the product of
a two-form and an eight-form.  
This is because the one-loop anomaly can only be canceled through
an additional anomalous variation of the classical action
which is necessarily so-factorized.  The reason why the 
classical contribution is so-factorized is because this contribution arises
only through the noninvariance of independent factors in the ``Chern-Simon's" 
interactions $CGG$ and $GX_7$.  But $SO(9,1)$ does not enable 
factorization of ${\rm tr}\,R^6$ (which is therefore said
to be the irreducible part of the anomaly),
so this term must vanish
identically. Therefore the dimension of the group ${\cal G}_i$
must be $n_i=248$.  Without yet specifying which 248-dimensional
group is permitted, we substitute 248 for $n_i$, obtaining
\brr I_{12}({\rm 1\,loop})_i &=& 
     \frac{1}{2\,(2\pi)^5\,6!}
     \bpl\,-\ft{15}{16}\,{\rm tr}R^4\,{\rm tr}R^2
     -\ft{15}{64}\,({\rm tr}R^2)^3
     +\ft{1}{16}\,{\rm tr}R^4\,{\rm Tr}\,F^2_i
     \nonumber\\[.1in]
     & & \hspace{1in}
     +\ft{5}{64}\,({\rm tr}R^2)^2\,{\rm Tr}\,F^2_i
     -\ft{5}{8}{\rm tr}R^2\,{\rm Tr}\,F^4_i
     +{\rm Tr}\,F^6_i\,\bpr \,,
\label{hwloop}\err
where factors ${\rm Tr}\,F_i^2$ involve a trace over the adjoint
representation of ${\cal G}_i$.
As explained above, it is necessary that this polynomial factorize
as the product of a two-form and an eight-form.  
It is straightforward
to algebraically impose this restriction, from which the following
is found to be a necessary requirement,
\brr {\rm Tr}F^6_i=
    \ft{1}{24}\,{\rm Tr}F^4_i\,{\rm Tr}F^2_i
    -\ft{1}{3600}\,({\rm Tr}F^2_i)^3 \,.
\label{property11}\err
There is exactly one nonabelian Lie group with this property,
$E_8$.
Given the property (\ref{property11}), as well as the
conventional $E_8$ definition of a $``{\rm tr}"$ operation,
${\rm Tr}\,F^2\equiv 30\,{\rm tr}\,F^2$, and the two other $E_8$ 
identities listed in appendix A,  after a small amount of straightforward
algebra the anomaly polynomial (\ref{hwloop}) can be reexpressed as follows,
\brr I_{12}({\rm 1\,loop})_i=\ft13\,\pi\,I_{4\,(i)}^{\,3}
     +X_8\wedge I_{4\,(i)} \,,
\label{fachw}\err
where $X_8$ is the eight-form given in (\ref{x8def}) and
$I_{4\,(i)}$ is a four-form given by
\brr I_{4\,(i)}=\frac{1}{16\pi^2}\,(\,
    {\rm tr}\,F_i^2-\ft12\,{\rm tr}R^2\,)\,.
\label{i4def}\err
The factorization (\ref{fachw}) was first presented in \cite{hw1}, and
provides the key to anomaly cancellation.
The first term of (\ref{fachw}) is canceled by 
inflow mediated by the $CGG$ interaction while the second term
is canceled by inflow mediated by the $GX_7$ interaction.

As in the case of the fivebrane described in the previous section,
the one-loop anomaly (\ref{hwloop}) is cancelled via inflow from
classical variation of the eleven-dimensional action.
This can include an anomalous contribution 
localized on the two ten-planes provided
the Bianchi identity $dG$ is appropriately modified. 
Anomaly inflow then arises due to the variation of the same two terms   
(\ref{action}) which were instrumental to the fivebrane 
anomaly cancellation.  A remarkable feature of {\it M}-theory is that
for this orbifold, as well as the one presented in the next section,
anomaly cancellation does not require any additional counterterms.

\vspace{.1in}
\noindent
{\it Magnetic couplings:}\\[.1in]
The modifications to the $dG$ Bianchi identity which enables the
necessary inflow mechanism can be derived in a systematic way.
Since $dG$ is a five-form, the most general modification with
local support on the ten-plane $M^{10}_i$ would have to be a gauge-invariant
four-form wedged with the one-form brane current $\d^{(1)}_{M^{10}_i}$.
The most general gauge-invariant four-form which is available for
this purpose would be some linear
combination of ${\rm tr}\,R^2$ and ${\rm tr}\,F_i^2$.  
In a fully general analysis, we would leave the coefficients of these 
two terms unspecified, finding later that they are fixed by anomaly
cancellation.  Not suprisingly it is precisely the combination
$I_{4\,(i)}$ given in (\ref{i4def}) which is required.  
For the sake of economy we sacrifice a very
small amount of ultimately irrelevant generality by specializing 
to this case. So the appropriately modified Bianchi identity is given by
\brr dG=\sum_{i=1}^2 I_{4\,(i)}\wedge\d^{\,(1)}_{M^{10}_i} \,,
\label{bianchihw}\err
where $I_{4\,(i)}$ is the four-form given in (\ref{i4def}) and
$\d^{\,(1)}_{M^{10}_i}$ is a one-form brane-current with
support on the $i$th ten-plane. 

\vspace{.1in}
\noindent
{\it Anomaly inflow:}\\[.1in]
To determine how the $CGG$ term
transforms, we need to determine how the three-form potential $C$
transforms.  To determine this, we need an explicit form for $G$.
This is determined as the object whose exterior
derivative reproduces the right-hand side of (\ref{bianchihw}).
This implies
\brr G &=& dC
    +\sum_{i=1}^2\,\bpl\,(b-1)\,\d_{M^{10}_i}^{\,(1)}\wedge\w_{3\,(i)}^0
    +\ft12\,b\,\theta_{(i)}\,I_{4\,(i)}\,\bpr \,,
\label{gsolvehw}\err
where $\w_{3\,(i)}^0$ is the Chern-Simons three-form determined by
$d\w_{3\,(i)}^0=I_{4\,(i)}$, while $\theta_{(i)}$ is
a zero-form with the two properties 
$d\theta_{(i)}=2\d^{(1)}_{M^{10}_i}$ and $\theta_{(i)}^2=1$,
and $b$ is a real parameter unspecified by the Bianchi identity.  
This parameter is, however, fixed by anomaly cancellation, as described below.

Since the field strength $G$ must be gauge invariant, this requires
that $C$ have the following transformation property under
gauge transformations and local Lorentz transformations,
\brr \d C=\sum_{i=1}^2\,(b-1)\,\w_{2\,(i)}^1\wedge\d_{M^{10}_i}^{\,(1)} \,.
\label{dchw}\err
Unless $b=1$, equation (\ref{dchw}) implies that $C$ has a nontrivial
transformation rule.  
In fact, as will be shown, anomaly cancellation {\it requires} $b=2$,
so this allows for anomaly inflow through the resulting noninvariance of
the $CGG$ interaction.  

Using the properties of the brane-currents described in section 2,
it is then straightforward to determine the transformation of
the two interactions $CGG$ and $GX_7$.  For the case of the
$CGG$ interaction we determine
\brr \d(\,-\frac{\pi}{3}\,\int C\wedge G\wedge G\,)
     &=& -\frac{\pi}{3}\,\sum_{i=1}^2\, \ft14\,(b-1)\,b^2\,\int_{M^{10}_i}\,
     \w_{2\,(i)}^1\wedge I_{4\,(i)}\wedge I_{4\,(i)}  \,.
\label{cgghw}\err
To obtain this result, we note that, since $G$ is gauge invariant, 
only the variation 
of the factor $C$ on the left-hand side of (\ref{cgghw}) contributes.
Using the explicit result (\ref{dchw}), this tells us that
$\d\int CG^{\,2}=(b-1)\sum_i\int_{M^{10}_i} \w_{2\,(i)}^1(G^{\,2}\,|)$, where 
the bar indicates
that $G^{\,2}$ is evaluated on the $i$th fixed ten-plane. Since 
$C_{ABC}|=0$, only the terms $\der_{[A}C_{BC](11)}$ contribute to
$dC|$, so that $dC|$ necessarily includes a $dx^{11}$ factor. 
Since both $\d C$ and $\d^{(1)}_{M^{10}_i}$ also contain $dx^{11}$ factors, 
we can therefore neglect the first two terms on the right hand side of
(\ref{gsolvehw}) when evaluating $G^{\,2}|$ 
(because $dx^{11}\wedge dx^{11}=0$). 
As a result, $G^{\,2}|$ is
proportional to $\t_{(i)}^{\,2}=1$ so that the product $G^{\,2}$ is 
well-defined on $M^{10}_i$.
Equation (\ref{cgghw}) describes anomaly inflow to the two fixed 
ten-planes due to the $CGG$ interaction.  

Similarly, we determine
\brr \d(\,\int G\wedge X_7) 
     = -\sum_{i=1}^{2}\,\int_{M^{10}_i}\,I_{4\,(i)}\wedge X_6^1\,.
\label{gx7hw}\err
To obtain (\ref{gx7hw}) ,
we have integrated by parts and used the Bianchi identity (\ref{bianchihw}).  
Equation (\ref{gx7hw}) describes inflow to the two
ten-planes due to the $GX_7$ interaction.  

The anomaly inflow can be described by a pair of twelve-forms 
$I_{12}({\rm inflow})_i$ which give rise to
(\ref{cgghw}) and (\ref{gx7hw}) upon descent.
Thus,
\brr I_{12}(CGG)_i &=& -\frac{\pi}{12}\,(b-1)\,b^2\,I_{4\,(i)}^{\,3} 
     \nonumber\\[.1in]
     I_{12}(GX_7)_i &=& -I_{4\,(i)}\wedge X_8 \,.
\label{inflowhw}\err
These two contributions must conspire to cancel against the quantum 
anomaly given in (\ref{fachw}).

\vspace{.1in}
\noindent
{\it The total anomaly:}\\[.1in]
The total anomaly is given by the sum of the
one-loop anomaly (\ref{fachw}) and the inflow contributions
(\ref{inflowhw}). Thus, 
$I_{12}({\rm total})=I_{12}({\rm 1\,loop})+I_{12}(CGG)+I_{12}(GX_7)$.
We require that this total anomaly vanish.  Nicely, the second term
of (\ref{fachw}) is canceled by $I_{12}(GX_7)_i$.
The first term
of (\ref{fachw}) is canceled by $I_{12}(CGG)_i$ provided $b$
satisfies the cubic equation $b^3-b^2-4=0$.  
This equation has one real root,
so anomaly cancellation uniquely selects 
\brr b=2 \,.
\err  
This value of $b$ is fixed by the consistency requirements. 
It is gratifying that this requirement is satisfied by a 
rational (indeed, integer) value for $b$. 
In contrast to the fivebrane worldvolume anomaly, 
the $CGG$ term provides a crucial ingredient to the removal of 
the gravitational anomaly.

\vspace{.1in}
\noindent
{\it Comment 1:}\\[.1in]
The above analysis is easily generalized to the case where fivebranes
propagate in the ${\bf R^{10}}\times S^1/{\bf Z}_2$ background. 
If there are $N_5$ fivebranes, then (\ref{bianchihw}) is
replaced with
\brr dG=\sum_{i=1}^2 I_{4\,(i)}\wedge\d^{\,(1)}_{M^{10}_i}
     +\sum_{i=1}^{N_5}\d^{(5)}_{W^6_i} \,,
\label{bianchihwfb}\err
where $W^6_i$ is the worldvolume of the $i$th fivebrane.
In this case, consistency is automatically assured because
any additional anomalies associated with the fivebranes are removed by 
the mechanism described in section 2.

\vspace{.1in}
\noindent
{\it Comment 2:}\\[.1in]
An essential point is that the factorization criterion
(\ref{property11}) differs
from the analogous criterion encountered in the 
effective field theory describing the perturbative 
heterotic string, obtained as the 
limit that the eleventh dimension shrinks to zero size.
In that limit, the theory becomes ten-dimensional and the
two orbifold ten-planes coincide.
The quantum anomaly is then replaced by the
{\it sum} of the two previously independent polynomials 
$I_{12}({\rm 1\,loop})_1$ and $I_{12}({\rm 1\,loop})_2$
since these are now
evaluated on the same manifold.  Furthermore, the D=11
supergravity theory collapses to D=10 N=1 supergravity and
the inflow mechanism involving the $CGG$ and the $GX_7$ terms
mutates into the ordinary Green-Schwarz mechanism mediated by the 
surviving two-form, which {\it also} requires factorization 
of the quantum anomaly polynomial.  However, since the
quantum anomaly in the small-radius limit 
is given by the {\it sum} of the two polynomials described above,
the factorization criterion is not the same. 
(The ten-dimensional requirement is derived in appendix B and
given as (\ref{property}), which should be compared with 
(\ref{property11})).
The ten-dimensional condition does allow
$E_8\times E_8$ as an allowed gauge group, as one would expect,
but it also has another solution, $SO(32)$, which is not
relevant to the full {\it M}-theory construction.
The fact that $SO(32)$ is found in the small-radius limit
but not on the expanded orbifold
can be phrased as an inability to ``pull apart" the 
gauge group $SO(32)$ to enable anomaly
cancellation on independent orbifold planes.  The gauge group
$E_8\times E_8$, on the other hand, does have the ability to
be ``pulled apart" so that each $E_8$ factor can be naturally 
associated with one of the two fixed ten-planes in the ${\bf Z}_2$ 
orbifold. The concept of ``pulled apart" twisted matter illustrates one
difference between what we refer to as ``collective" anomaly
constraints compared to ``local" anomaly constraints, the former
referring to weaker conditions which apply in the collapsed
limit when orbifold planes coelesce.

\section{{\it M}-theory on ${\bf R}^6\times T^5/{\bf Z}_2$}
There is another {\it M}-theory orbifold which gave rise, through a
detailed analysis of local anomaly cancellation, to 
yet another important
insight into a subtle mechanism of {\it M}-theory.  In this case
D=11 supergravity on ${\bf R}^6\times T^5/{\bf Z}_2$,
which was first analyzed in \cite{dasmuk, wittens5}, 
a curious degeneracy first implicated the role of ``wandering fivebranes"
in mediating phase transitions.

As in the case of the $S^1/{\bf Z}_2$ orbifold described in the 
previous section, the $T^5/{\bf Z}_2$ orbifold has a set of parallel
fixed-planes which are indistinguishable aside from their position.
But in the $T^5/{\bf Z}_2$ case, there are thirty-two fixed-planes 
rather than two,
and their dimensionality is six rather than ten.
Due to important differences from the previous case, and also
because much of the analysis in the more complicated orbifolds 
with intersecting fixed planes again parallels this discussion,
it is worthwhile to briefly but completely review
the anomaly analysis for the case of the $T^5/{\bf Z}_2$ orbifold.

\vspace{.1in}
\noindent
{\it Structure of the orbifold:}\\[.1in]
Start with eleven-dimensional supergravity on ${\bf R}^6\times T^5$.  The
${\bf R}^6$ factor is parameterized by $x^\mu\equiv\{x^1,...,x^6\}$, while
the five compact dimensions are parameterized by 
$x^i\equiv \{x^7,...,x^{11}\}$, which  
each takes values on the interval $[-\pi,\pi]$ with endpoints identified.
Truncate the theory via a ${\bf Z}_2$ projection which
acts on each of the five compact coordinates
as $x^i\rightarrow -x^i$. There are 
$2^5=32$ six-dimensional hyperplanes fixed by this projection,
namely the surfaces defined when each of the five 
$x^i$ independently assumes the value 0 or $\pi$.

Demanding ${\bf Z}_2$ invariance of the eleven-dimensional $CGG$ 
interaction term
implies that the three-form $C$ is odd under
the ${\bf Z}_2$ projection. 
Thus, $C_{\mu\nu\rho}\rightarrow -C_{\mu\nu\rho}$
and $C_{\mu ij}\rightarrow -C_{\mu ij }$.
The components $C_{\mu\nu\rho}$ and $C_{\mu ij}$ therefore vanish on the 
fixed six-planes.  
However, $C_{\mu\nu i}\rightarrow C_{\mu\nu i}$.  
Therefore, from the point of view of the six-dimensional fixed-planes,
the three-form $C$ contributes {\it five} nonvanishing tensors
$C_{i\mu\nu}$.  

Half the supersymmetries of the eleven-dimensional theory are broken
on the fixed-planes
by the ${\bf Z}_2$ projection.  Thus, on the fixed-planes
the five six-dimensional two-forms $C_{i\mu\nu}$ organize
along with the other fields surviving the ${\bf Z}_2$ projection
into D=6 N=2 supermultiplets. Since there are no 
six-dimensional vector fields surviving the ${\bf Z}_2$ projection, it
follows that the relevant supersymmetry is the chiral
N=2b theory. This is because the alternative, the nonchiral
N=2a theory, necessarily involves vector fields.  The N=2b
supergravity multiplet includes five self-dual two-forms, so   
the remaining five anti self-dual components must 
organize into matter supermultiplets. 
In the N=2b theory, the only matter multiplet 
is the tensor multiplet (which includes one anti self-dual two-form).  
Thus, in addition to the N=2b supergravity, we have 
five D=6 N=2 tensor multiplets in the ``untwisted sector".
The anomaly due to the self-dual and anti self-dual two-forms 
cancel each other.  So the anomalous ``untwisted" couplings involve
two chiral spin 3/2 fields coming from the supergravity 
multiplet and ten antichiral spin 1/2 fields coming two each from the 
five tensor multiplets.

We also allow for a ``twisted sector" involving some number 
$n_i$ of D=6 N=2 tensor multiplets to propagate on the $i$th six-plane.
The anomalous ``twisted" couplings therefore involve $n_i$ anti self-dual
tensor fields and $2 n_i$ antichiral spin 1/2 fields.

\vspace{.1in}
\noindent
{\it The one-loop anomaly:}\\[.1in]
A gravitational anomaly arises due to the coupling 
of chiral projections of the bulk gravitino to currents
localized on the thirty-two fixed-planes.  
Since the six-planes are indistinguisable
aside from their position, this anomaly is similar on each
of the thirty-two, and can be computed using standard formulae if
proper care is used.  The reason why extra care is needed 
is that the the anomaly in question actually derives from the coupling of 
{\it eleven}-dimensional fields to {\it six}-dimensional currents,
whereby standard index theorem results (such as those described in
appendix C) can be applied directly only in the small-radius
limit when the thirty-two fixed-planes coincide.  This is because it
is only in this limit that we actually describe
six-dimensional fields coupled to six-dimensional currents.  
Thus, we can compute the gravitino-induced 
anomaly on a given six-plane as 
1/32 of that described by the index theorem results 
using the (six-dimensional) untwisted spectrum described above.

The ``twisted" tensor multiplets also contribute a gravitational anomaly.
However, since the twisted fields are
six-dimensional these can be computed directly using the standard formulae
(without multiplying by 1/32).

It is then straightforward to compute the one-loop anomaly using
the formulae described in appendix C.  
The local anomaly on the $i$th six-plane is characterized 
by the following eight-form,
\brr I_8({\rm 1\,loop})_i &=& \frac{1}{32}\bpl\,
     2\,I_{\rm GRAV}^{(3/2)}(R)
     -10\,I^{(1/2)}_{\rm GRAV}(R)\,\bpr
     \nonumber\\[.1in]
      & & -n_i\,\bpl\,2I^{(1/2)}_{\rm GRAV}(R)
     +I_{\rm GRAV}^{({\rm 3-form})}(R)\,\bpr \,,
\label{s52}\err
where the various constituent polynomials are
given explicitly in (\ref{grav6}), and
$n_i$ is the number of twisted tensor multiplets.
The first line describes the untwisted anomaly
and includes an additional factor of 1/32 
for the reasons described in the previous paragraph.
The second line describes the twisted anomaly coming from
$n_i$ tensor multiplets.

We compute the anomaly by substituting the explicit polynomials 
given in (\ref{grav6}) into the expression (\ref{s52}).  
After a small amount of algebra the result organizes as 
\brr I_8({\rm 1\,loop})_i=(n_i-\ft12)\,X_8 \,,
\label{loopt5}\err
where $X_8$ is given in (\ref{x8def}).  Since the number of tensor 
multiplets should be integer, it is apparent 
that another mechanism is required to cancel this anomaly. 
In fact, as in the case of the fivebrane described in section 2
and also in the $S^1/{\bf Z}_2$ orbifold described in section 3,
the one-loop anomaly (\ref{loopt5}) is canceled via inflow
from classical variation of the eleven-dimensional action.
This can include an anomalous contribution localized on the
thirty-two fixed planes provided the Bianchi identity $dG$
is appropriately modified.
Anomaly inflow then arises due to the variation of the $GX_7$ term.

\vspace{.1in}
\noindent
{\it Magnetic couplings:}\\[.1in]
Since $dG$ is a five-form, the most general modification with
local support on the six-plane $M^6_i$ would have to be a zero-form
(ie: a number) multiplying the five-form brane current $\d^{(5)}_{M^6_i}$.
So the appropriately modified Bianchi identity is given by
\brr dG=\sum_{i=1}^{32} g_i\,\d^{\,(5)}_{M^6_i} \,,
\label{bianchit5}\err
where $\d^{\,(5)}_{M^6_i}$ is a five-form brane-current with
support on the $i$th six-plane $M^6_i$ and $g_i$ are
yet-unspecified rational magnetic charges assigned independently
to each of the $M^6_i$.

\vspace{.1in}
\noindent
{\it Anomaly inflow:}\\[.1in]
It is straightforward to determine the transformation of
the $GX_7$ interaction.  This is found to be
\brr \d(\,\int G\wedge X_7) 
     = -\sum_{i=1}^{32}g_i\,\int_{M^6_i}\,X_6^1\,.
\label{gx7t5}\err
To obtain (\ref{gx7t5}) 
we have integrated by parts and used the Bianchi identity (\ref{bianchit5}).  
Equation (\ref{gx7t5}) describes inflow to the 
six-dimensional fixed planes due to the $GX_7$ interaction.  
Note that in this case the $CGG$ interaction does not
contribute inflow to the local gravitational anomaly.

The anomaly inflow is characterized by the eight-form
$I_8(GX_7)_i$ which gives rise to
(\ref{gx7t5}) upon descent.  Thus,
\brr I_8(GX_7)_i &=& -g_i\,X_8 \,,
\label{inflowt5}\err
This contribution must conspire to cancel against the quantum anomaly given 
in (\ref{loopt5}).

\vspace{.1in}
\noindent
{\it The total anomaly:}\\[.1in]
The total anomaly is given by the sum of the
one-loop contribution (\ref{loopt5}) and the inflow contribution
(\ref{inflowt5}). Thus, 
$I_8({\rm total})_i=I_8({\rm 1\,loop})_i+I_8(GX_7)_i$, which tells us
\brr I_8({\rm total})_i=(n_i-g_i-\ft12)\,X_8\,.
\err
We require that this total anomaly vanish.  
This requires that 
\brr n_i-g_i=\ft12 \,.
\label{ng}\err
Note that whereas $n_i$ is necessarily an integer, the magnetic
charge $g_i$ can be half-integer.  The anomaly vanishes for
a given choice of magnetic charge $g_i$ provided there are
$n_i=g_i+1/2$ tensor multiplets.  Since $n_i$ should be a 
nonnegative integer this tells us that there is a minimum
magnetic charge equal to $-1/2$ with permissible values at successively 
greater integer increments. Thus, the allowed values of $g_i$ are given by
$-1/2,\,+1/2,\,+3/2,...$.

\vspace{.1in}
\noindent
{\it Comment 1:}\\[.1in]
The above analysis is easily generalized to the case where fivebranes
propagate in the $T^5/{\bf Z}_2$ background. 
If there are $N_5$ fivebranes, then (\ref{bianchit5}) is
replaced with
\brr dG=\sum_{i=1}^{32} g_i\,\d^{\,(5)}_{M^6_i}
     +\sum_{i=1}^{N_5}\d^{(5)}_{W^6_i} \,,
\label{bianchit5fb}\err
where $W^6_i$ is the worldvolume of the $i$th fivebrane.
In this case, consistency is automatically assured because
any additional anomalies associated with the fivebranes are removed by 
the mechanism described in section 2.

\vspace{.1in}
\noindent
{\it Global constraints and wandering fivebranes:}\\[.1in]
If we integrate the Bianchi identity (\ref{bianchit5fb}) over 
the compact $T^5$, the left-hand side vanishes due to Stokes
theorem (since there is no boundary) and the right-hand
becomes $N_5+\sum_ig_i$ (since the brane-currents integrate to unity).  
Therefore 
\brr N_5+\sum_i g_i=0 \,.
\label{n5g}\err
Now, if we sum equation (\ref{ng}) over the 32 fixed points using this 
constraint, we determine a second global constraint given by
$N_5+\sum_i n_i=16$.
Thus, the total number of twisted tensor multiplets plus the number
of fivebranes must be 16.  Since each fivebrane also supports
a tensor multiplet, this tells us that we have a total of 16 
tensor multiplets. 
There are various ways to realize all of these constraints.
For instance, if there are no fivebranes (so that $N_5=0$), 
one could place tensor multiplets on 16 of the 32 fixed points, 
and assign
magnetic charge $+1/2$ to these same 16 fixed points and magnetic charge
$-1/2$ to the remaining 16 fixed points which do not support 
tensor multiplets.   There are numerous other possibilities.
However, there is no way to solve all of the constraints 
in a way which treats all 32 fixed-planes identically unless 
we allow $N_5\ne 0$.

The most symmetrical individual solution has $N_5=16$
and identical magnetic charge $g_i=-1/2$ for each of the
thirty-two fixed planes.
These assignments satisfy the global constraint
(\ref{n5g}).  Since $g_i=-1/2$, the local
constraint (\ref{ng}) requires that $n_i=0$, so that there are 
no twisted tensor multiplets in this solution.
 
The presence of fivebranes also allows for a unified description
(first presented in \cite{wittens5}) which symmetrically encorporates
all of the distinct vacua.
This is obtained if all twisted tensors are associated with fivebranes
wrapping the fixed six-plane in question.  A picture emerges
by which fivebranes can detach from a given six-plane, taking one tensor
and one unit of charge with it, and ``wander" throughout the bulk.  
Similarly, a ``wandering fivebrane" can move to and wrap a particular 
six-plane, thereby adding one tensor to the twisted spectrum of that
plane and simultaneously increasing the magnetic charge by one.  
The wandering branes will have their anomalies canceled by the mechanism
explained in section 2, whereas the six-planes will have any 
local anomaly canceled by the similar mechanism explained above in 
this section.  In this way all of the unique non-symmetrical configurations
are linked by phase transitions mediated by the fivebranes!

\vspace{.1in}
\noindent
{\it Comment:}\\[.1in]
One may ponder another mechanism whereby, on a fixed six-plane,
a twisted tensor field mediates a Green-Schwarz mechanism locally
via counterterms in the action describing the tensor dynamics.  
For the case of the $T^5/{\bf Z}_2$ orbifold, however,
this isn't possible.  This is because the anomaly eight-form
is proportional to $X_8$, which is not factorizable
due to the presence of ${\rm tr}\,R^4$. In the case of the orbifold
presented in section 6, however, we find a scenario
where we have six-dimensional orbifold planes which can support
tensors, but where the anomaly can indeed factorize.  In that case,
such a local Green-Schwarz mechanism is not only possible,
but necessary.

\setcounter{equation}{0}
\section{The ``I-brane effect"}
The orbifolds analyzed in sections 3 and 4 are the simplest 
possible examples.  In those cases, all of the orbifold planes
are of the same dimensionality (ten in the $S^1/{\bf Z}_2$, and
six in the $T^5/{\bf Z}_2$ case) and do not intersect each other.
More generally oribifolds will involve fixed planes of
mixed dimensionalities which can intersect.  In this case, the
analysis of local anomaly cancellation is considerably more involved.
The presence of intersections allows for a type of inflow mechanism
not encountered in the simpler examples.  In this section, we
describe this effect in generality in a self-contained manner.  
In the next section, we analyze a ${\bf Z}_2\times {\bf Z}_2$
orbifold which involves this effect as a necessary ingredient.

Brane physics has spawned several cousin-effects to the 
the Green-Schwarz mechanism.  These are necessary to explain anomaly 
cancellation in various scenarios involving spacetime defects.
Whereby the conventional Green-Schwarz mechanism involves nontrivial 
magnetic and also nontrivial electric couplings to the two-form potential 
in ten-dimensional N=1 supergravity, neither of which is concentrated on
spacetime defects, the cousin-effects include couplings which are 
localized either on the worldvolume of branes or on orbifold
points.  In the conventional mechanism, a ten-dimensional 
quantum anomaly is cancelled by
a nonvanishing variation of the classical action involving an interplay 
between ten-dimensional magnetic couplings, which appear in the 
$dH$ Bianchi identity, and ten-dimensional
electric couplings,
which appear as Chern-Simons interactions.  
By way of comparison, the gravitational anomaly
of {\it M}-fivebranes is removed by an interplay between a 
{\it six}-dimensional magnetic coupling to the 
three-form potential of eleven-dimensional supergravity which is localized 
on the worldvolume of the fivebrane, and an {\it eleven}-dimensional
electric coupling appearing as a Chern-Simons interaction.  
In this section, we describe a related effect
inspired by considerations of intersecting D-branes, dubbed 
``I-branes" and first presented in \cite{ghm},
and generalized and nicely explained in \cite{cy}.

The idea behind the ``I-brane effect" which we consider involves
magnetic and electric couplings to the three-form potential 
which are variously supported on 
submanifolds.  For instance, a magnetic coupling 
(appearing in the $dG$ Bianchi identity) can be
concentrated on a $d1$-dimensional defect.
Additionally, electric couplings can appear as Chern-Simon's
interactions in the worldvolume Lagrangian describing 
matter propagating on a $d2$-dimensional defect.  
This could be ``twisted" matter concentrated on an orbifold plane.  
The electric couplings would then  
appear also in the classical field equation for $G$ obtained by the
variational principle.  Thus, the magnetic and electric couplings
appear in the Bianchi identity and in the classical field 
equations, respectively, as follows
\brr d G &=& 
     \d_{M^{d1}}^{(11-d1)}\wedge\tilde{Y}_{d1-6}
     \nonumber\\[.1in]
     d\star G &\propto& 
     \d_{M^{d2}}^{(11-d2)}\wedge Y_{d2-3} \,,
\label{ibcouplings}\err 
where $\tilde{Y}_{d1-6}$ is a closed, gauge-invariant 
$(d1-6)$-form coupling magnetically
to $G$, while $Y_{d2-3}$ is a closed, gauge-invariant 
$(d2-3)$-form coupling electrically to $G$.  The forms
$\d_{M^{d}}^{(11-d)}$ are $(11-d)$-form brane-currents.

The second equation in (\ref{ibcouplings}) derives from a
Chern-Simon's interaction localized on the submanifold $M^{d2}$,
given by
\brr S_{CS}(M^{d2})=\int\d_{M^{d2}}^{(11-d2)}\wedge G\wedge
     Y^0_{d2-4} \,,
\label{d2cs}\err
where $Y_{d2-3}=d\,Y^0_{d2-4}$.  This is easily verified, as 
the classical variation $\d/\d G$ applied to (\ref{d2cs}) 
gives the the right-hand side of the second equation of 
(\ref{ibcouplings}), whereas variation of the 
$G\star G$ kinetic term supplies the left-hand side.

Varying (\ref{d2cs}), integrating by parts, and using
the first line of (\ref{ibcouplings}), it is straightforward 
to show that
\footnote{We use a standard notation to describe forms linked by descent,
such that a closed gauge invariant form $Z_q$ is written locally as
$Z_q=dZ^0_{q-1}$, where $Z^0_{q-1}$ has gauge variation
$\d Z^0=dZ^1_{q-2}$.}
\brr \d S_{CS}(M^{d2}) &=& -\int\d^{(11-{\cal I})}_{M^{d2}\cap M^{d1}}\wedge
     \bpl\,\tilde{Y}_{d1-6}\wedge Y_{d2-3}\,\bpr^1_{{\cal I}} 
     \nonumber\\[.1in]
     &=& -\int_{M^{\cal I}}
     \bpl\,\tilde{Y}_{d1-6}\wedge Y_{d2-3}\,\bpr^1_{{\cal I}} \,\,\,,
\label{iba}\err
where ${\cal I}=d1+d2-11$ is the dimensionality of the
intersection $M^{\cal I}=M^{d1}\cap M^{d2}$.
Thus, there is an anomalous classical variation localized on the
${\cal I}$-dimensional intersection of $M^{d1}$ and $M^{d2}$. 
This anomaly is characterized by the $({\cal I}+2)$-form which gives 
rise to (\ref{iba}) upon descent.  Thus,
\brr Y(IB)_{{\cal I}+2}=\tilde{Y}_{d1-6}\wedge Y_{d2-3} \,.
\label{yib}\err
  
The intersection anomaly (\ref{yib}) involes
the product of a magnetically-coupled form localized on one defect 
with an electrically-coupled form localized on another defect, while
the contribution to the anomaly itself is localized on the intersection.

\setcounter{equation}{0}
\section{{\it M}-Theory on $S^1/{\bf Z}_2\times T^4/{\bf Z}_2$}
In this section, we consider the simplest nontrivial {\it M}-theory
orbifold which involves multiple intersecting fixed-planes.
This example has fixed planes of ten-, seven- and six-dimensions,
the six-planes lying at the intersections of the ten-planes
with the seven-planes.  It is actually a second ${\bf Z}_2$ orbifolding 
of the ``Ho{\v r}ava-Witten" ${\bf Z}_2$ orbifold described in 
section 3, and represents a 
singular limit of {\it M}-theory on $S^1/{\bf Z}_2\times K3$.

The greater complexity of this orbifold compared to the
${\bf Z}_2$ orbifolds necessitates a greater systematics.  
We first define more precisely the structure of the orbifold,
then develop the needed machinery, and then use this to determine
the twisted states.  It turns out that cancelation
of the anomaly at six-dimensional orbifold-plane intersections requires
particular twisted states on the entirety of one of the intersecting
planes, which is seven-dimensional.  This analysis illustrates how
gravitational anomalies can be used to determine states 
in extended regions without a continous local anomaly.

\vspace{.1in}
\noindent
{\it The structure of the orbifold:}\\[.1in]
We consider the specific $S^1/{\bf Z}_2\times T^4/{\bf Z}_2$ 
orbifold defined as follows.
Start with eleven-dimensional supergravity on a spacetime
with topology ${\bf R}^6\times T^5$.
The five compact coordinates, $\{x^7,x^8,x^9,x^{10},x^{11}\}$
each takes values on the interval $[-\pi,\pi]$ with endpoints
identified.  In addition to the unit element, the orbifold group 
includes an element $\a$ which reverses the orientation of the 
eleventh coordinate, $x^{11}\!\rightarrow\!-x^{11}$, an element $\b$ 
which reverses the orientation on each of the
four coordinates $x^i\equiv\{x^7,x^8,x^9,x^{10}\}$,
and the product $\a\b$ which reverses the orientation of
all five compact coordinates.  The action of the three nontrivial 
elements are displayed in table 1.

\begin{figure}
\begin{center}
\begin{tabular}{|c|ccccc|}
\hline
&&&&& \\[-.1in]
& \hspace{.1in}$x^7$\hspace{.1in} & 
\hspace{.1in}$x^8$\hspace{.1in} & 
\hspace{.1in}$x^9$\hspace{.1in}& 
\hspace{.1in}$x^{10}$\hspace{.1in}& 
\hspace{.1in}$x^{11}$\hspace{.2in}\\[.1in]
\hline
&&&&& \\[-.1in]
$\a$   & + & + & + & + & $-$ \\[.1in]
$\b$   & $-$ & $-$ & $-$ & $-$ & + \\[.1in]
$\a\b$ & $-$ & $-$ & $-$ & $-$ & $-$ \\[.1in]
\hline
\end{tabular}\\[.2in]
\parbox{4in}{Table 1: The action of the orbifold group ${\bf Z}_2\times {\bf Z}_2$ on 
the five compact coordinates of the orbifold discussed in
section 6.  A plus sign 
indicates no action on the indicated coordinate and a minus sign
indicates a parity reversal, $x^i\rightarrow -x^i$.}
\end{center}
\end{figure}

The global structure of this orbifold is determined as follows.
The element $\a$ leaves invariant the two ten-planes
defined by $x^{11}=0$ and $x^{11}=\pi$, while
$\b$ leaves invariant the sixteen seven-planes defined 
when the four coordinates $x^i$ individually assume 
the value $0$ or $\pi$.  Finally, $\a\b$ leaves invariant 
the thirty-two six-planes defined when
all five compact coordinates individually assume the value $0$ or
$\pi$.  The $\a\b$ six-planes coincide with 
intersections of the $\a$ ten-planes with the 
$\b$ seven-planes. The global structure is nicely visualized by the
diagram in figure 1. 

\begin{figure}
\begin{center}
\includegraphics[width=6in,angle=0]{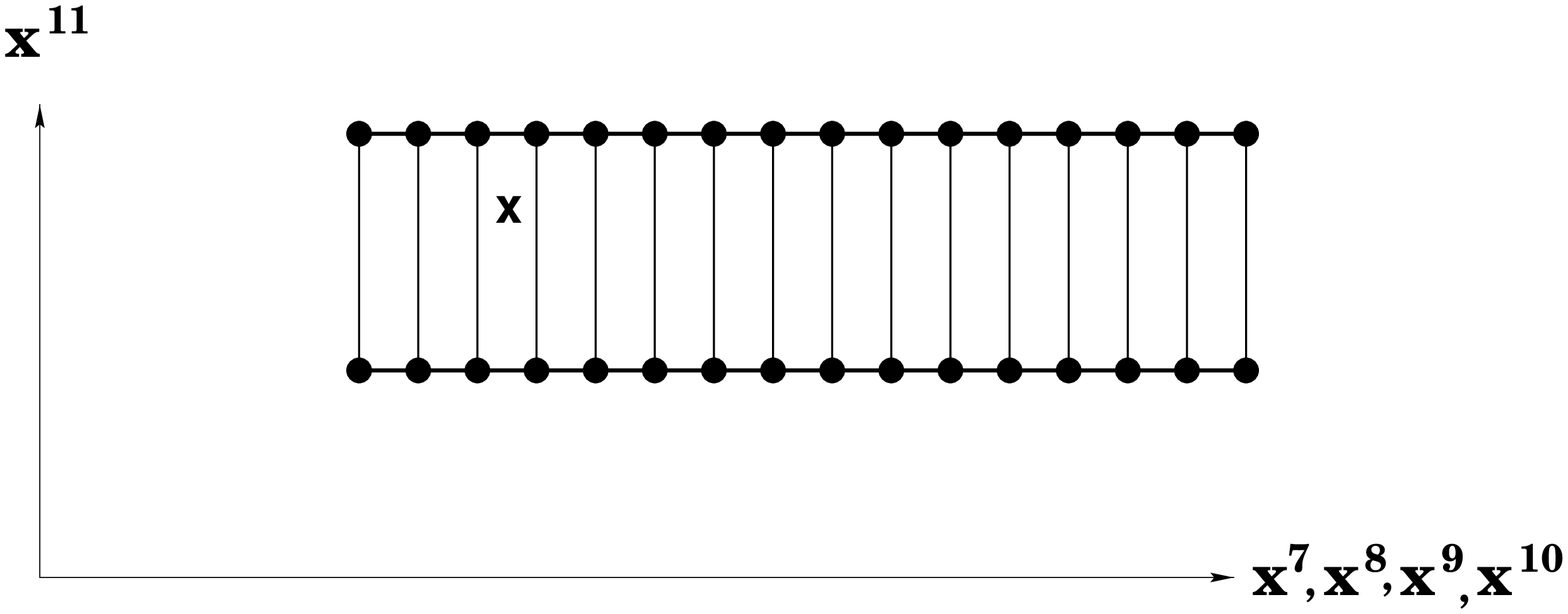}\\[.2in]
\parbox{6in}{Figure 1:  The global structure of orbifold planes
in the $S^1/{\bf Z}_2\times T^4/{\bf Z}_2$ orbifold.  
The two horizontal lines represent the
two ten-dimensional (``Ho{\v r}ava-Witten") fixed planes associated
with the ${\bf Z}_2$ factor denoted $\a$, while
the sixteen vertical lines represent the seven-dimensional
fixed-planes associated with the ${\bf Z}_2$ action denoted $\b$.
The thirty-two six-dimensional fixed planes assocated with
$\a\b$ are represented by the solid dots.  These coincide
with the intersection of the $\a$ planes and the $\b$ planes.
The $X$ in the figure indicates the presence of a ``wandering"
fivebrane as described in the text.}
\end{center}

\end{figure}

Thus, this orbifold describes a network 
six-, seven-, and ten-dimensional fixed-planes which
intersect.
As described above, chiral projections of bulk objects localized
on the fixed-planes induce anomalies at one loop.  Additional
contributions appear via ``inflow" from classical
variation of the bulk theory which may include anomalous 
pieces localized on defects.  These occur
when $G$ couples magnetically to the defects
via modififations to the $dG$ Bianchi identity,
and follow specifically due to the variation of the terms 
shown in (\ref{action}).

\vspace{.1in}
\noindent
{\it Magnetic and Electric sources for $G$:}\\[.1in]
The most general modified Bianchi identity will include terms
supported locally on orbifold planes and on fivebrane worldvolumes.
Since $dG$ is a five-form, a term with local support on one of
the seven-planes would have to be proportional to a closed,
gauge-invariant one-form constructed from the available fields,
wedged with the four-form brane-current $\d^{(4)}_{M^7_i}$.  But
there is no way to construct a closed, gauge-invariant one-form
from the available fields, so
this kind of coupling is disallowed.  Therefore, the most general
Bianchi identity is given by
\brr dG=\sum_{i=1}^2\,I_{4(i)}\,\d^{\,(1)}_{M^{10}_i}    
    +\sum_{i=1}^{32}\,g_i\,\d^{\,(5)}_{M^6_i}
    +\sum_{i=1}^{N_5}\,\d^{\,(5)}_{W^6_i}\,,
\label{bianchi1}\err
where $\d^{\,(1)}_{M^{10}_i}$ has support on 
the $i$th ten-plane $M^{10}_i$ while 
$\d^{\,(5)}_{M^6_i}$ and $\d^{\,(5)}_{W^i}$ have support on the
six-planes $M^6_i$ and on the fivebrane worldvolumes $W^6_i$, respectively.
The four-form $I_{4\,(i)}$ is determined by anomaly cancellation on 
$M^{10}_i$ precisely as described in section 3, and is
defined in equation(\ref{i4def}).
Finally, $g_i$ are magnetic charges assigned independently 
to each of the thirty-two fixed six-planes.

The seven-planes do not couple magnetically to
$G$ due to their odd-dimensionality.  Alternatively, 
they can couple electrically in the manner described
in section 5.  This electric coupling can provide an anomaly
at seven-plane/ten-plane intersections due to an 
interplay with the ten-dimensional magnetic coupling.  We discuss
this further below.

A significant part of our analysis relates to determining the
magnetic charges $g_i$. We will present
an argument supporting quarter-integer values for $g_i$ in the context of
the ${\bf Z}_2\times {\bf Z}_2$ orbifold.  
In addition, there is
a (global) restriction which
correlates the total orbifold charge $\sum_ig_i$ with other 
topolological data pertaining to gauge and gravitational instantons and the 
number of fivebranes, and
is obtained by integrating (\ref{bianchi1}) over
the five-cycle spanned by the five compact dimensions.

The quarter-integer charge quantization can be motivated by the
following arguments.  The orbifold in question represents
a singular deformed limit of $K3\times S^1/{\bf Z}_2$, 
where all of the $K3$ curvature is ``pinched" to be concentrated and
symmetrically distributed over the sixteen fixed seven-planes 
associated with the element $\b$ of the orbifold group
(these seven-planes are represented by the 
vertical lines in figure 1). 
If we start with the smooth $K3$,
we can represent the singular orbifold limit by 
the following deformation of the eleven-dimensional curvature,
\brr {\rm tr}\,R^2\rightarrow
     {\rm tr}\,{\cal R}^2+\pi^2\,\chi\,\sum_{i=1}^{16}\d^{\,(4)}_{M^7_i} \,,
\label{limit}\err
where 
$\d^{(4)}_{M^7_i}$ has support on the $i$th fixed seven-plane $M^7_i$, 
$\chi=24$ is the
Euler number of $K3$, and $\int_T{\rm tr}\,{\cal R}^2=0$, where 
the integration is over a four-cycle $T$.
Note that this is consistent with the requirement that
$\int_T{\rm tr}\,R^2=16\,\pi^2\,\chi$.  If we make the replacement
(\ref{limit}) in the first term on the right-hand side of
the Bianchi identity (\ref{bianchi1}), 
we generate new terms given by
\brr \frac{1}{16}(-\frac{1}{2})\,\chi\,
     \sum_{i=1}^2\,\sum_{j=1}^{16}\,\d^{(1)}_{M^{10_i}}
     \wedge\d^{(4)}_{M^7_j}
     =\sum_{i=1}^{32}\,(-\frac{\chi}{32})\,\d^{(5)}_{M^6_i}
\err 
where we have used equation (\ref{prop}).  
These new terms, however, are absorbed by a shift
$g_i\to g_i-\chi/32$. In this way we see that,
in the orbifold limit, the Euler character of the smooth $K3$ is
equally distributed as a magnetic charge 
$-\chi/32=-3/4$
at each of the 32 orbifold fixed six-planes.   This represents the
gravitational contribution to the magnetic charge.
Similar reasoning 
motivates that gauge instantons can yield only positive integer 
or positive half-integer
\footnote{The half-integer instanton contributions are associated with 
the ALE instantons classified by the Stieffel-Whitney class.} 
contributions to $g_i$, whereas a resident fivebrane should 
contribute only positive integer values.  Thus, a given six-plane
$M^6_i$ should have a minimum charge equal to $-3/4$ with permissible
values at successively greater half-integer increments.  Thus, the
allowed values of $g_i$ would be 
\brr g_i= -3/4,\,-1/4,\,+1/4,...
\err 
As described below, the same restriction
on permissible values of $g_i$ is found in an independent manner
by factorization requirements on the anomaly polynomial.  

An additional (global) constraint follows from integrating the Bianchi identity
(\ref{bianchi1}) over all five compact dimensions.
Since this region has no boundary,
the left-hand side of the integrated version of 
(\ref{bianchi1}) vanishes due to Stokes theorem, since the integrand
is a total derivative.  
Due to the properties of the brane-currents described in section 2,
the integrated right-hand side of (\ref{bianchi1})
reduces to the sum $n_1+n_2-\chi+N_5+\sum_i g_i$, where
$n_i$ are the instanton numbers associated with the bundle
$E_{8\,(i)}\rightarrow M^{10}_{\,i}$
and $\chi$ is the Euler number defined by 
$16\pi^2\chi=\int_T {\rm tr}\,R^2$.
The instanton numbers and
the Euler number occur since the brane-current in the first term of
(\ref{bianchi1}) collapses the integral to a four-cyle
integral over $T$.  So the generic constraint is
\brr n_1+n_2-\chi+N_5+\sum_ig_i=0 \,.
\label{cons1}\err
However, we can simplify this expression considerably in 
the orbifold case where,
as we have explained, all of
the local curvature which contributes to the Euler number 
is concentrated on the orbifold-planes, and is more properly absorbed
into the magnetic charges $g_i$.  So an explicit $\chi$ term should
not be included in (\ref{cons1}).  Similar concerns apply
to any ``zero-size" gauge instanton which is trapped on one of the
fixed six-planes.  Furthermore, any gauge
instanton on one of the ten-planes which is {\it not} trapped at an orbifold singularity should be continuously deformable into a fivebrane
by shrinking its size to zero.  Such a fivebrane can then 
detach and move into the bulk.  
Since the two situations are smoothly related on moduli space, we
expect an anomaly-free solution for a fivebrane to imply an anomaly-free
solution for the associated instanton (and vice-versa).  
It is simpler to work with fivebranes and, hence, we
omit the instanton terms $n_1$ and $n_2$ in equation (\ref{cons1}).
In the analysis to follow, we therefore replace (\ref{cons1})
with the minimal constraint
\brr N_5+\sum_{i=1}^{32}g_i=0 \,,
\label{cons2}\err
noting that $N_5$ should be a positive
integer.  We also keep in mind the quarter-integer
quantization of $g_i$.  

\vspace{.1in}
\noindent
{\it Anomaly Inflow:}\\[.1in]
As in the cases of the ${\bf Z}_2$ orbifolds described in sections
3 and 4, the cancellation of the local anomaly involves one-loop
contributions and also inflow contributions describing
transformations of the $CGG$ and $GX_7$ terms.
There is also another kind of inflow at work in this case
which can arise because of the fixed-plane intersections.
This is an ``I-brane" contribution of the sort described in section 5.
The $CGG$ and $GX_7$ contributions are easiest 
to analyze, so we begin by working these out.  Following this, we
independently analyze the ``I-brane" contribution before 
moving on to the one-loop anomaly.

To determine how the $CGG$ term
transforms, we need to determine how the three-form potential $C$
transforms.  To determine this, we need an explicit form for $G$.
This is determined as the object whose exterior
derivative reproduces the right-hand side of (\ref{bianchi1}).
Suppressing the fivebrane contributions\footnote{
We suppress the fivebrane contribution to $G$ because
this involves unnecessary complexity; the contributions localized on
the fivebranes do not affect the orbifold-plane anomalies and
furthermore, as described in section 2, the fivebrane anomalies are 
independently resolved.}, this implies
the following definition,
\brr G &=& dC
    +\sum_{i=1}^2\,\bpl\,(b-1)\,\d_{M^{10}_i}^{\,(1)}\wedge\w_{3\,(i)}^0
    +\ft12\,b\,\theta_{(i)}\,I_{4\,(i)}\,\bpr
    +\sum_{i=1}^{32}\,\ft12\,g_i\,\theta_{(i)}\,\d_{M^7_i}^{\,(4)}\,,
\label{gsolve}\err
where $\w_{3\,(i)}^0$ is the Chern-Simons three-form determined by
$d\w_{3\,(i)}^0=I_{4\,(i)}$, while $\theta_{(i)}$ is
a zero-form with the two properties 
$d\theta_{(i)}=2\d^{(1)}_{M^{10}_i}$ and $\theta_{(i)}^{\,2}=1$,
and $b$ is a real parameter unspecified by the Bianchi identity.  
This parameter is, however, fixed by anomaly cancellation, as described below.

Since the field strength $G$ must be gauge invariant, this requires
that $C$ have the following transformation property,
\brr \d C=\sum_{i=1}^2\,(b-1)\,\w_{2\,(i)}^1\wedge\d_{M^{10}_i}^{\,(1)} \,.
\label{dc}\err
Unless $b=1$, equation (\ref{dc}) implies that $C$ has a nontrivial
transformation rule.  
In fact, as verified below, anomaly cancellation 
{\it requires} $b=2$.
This enables anomaly inflow through the resulting noninvariance of
the $CGG$ interaction. 

Using the properties of the brane-currents described in section 2,
it is then straightforward to determine the transformation of
the two interactions $CGG$ and $GX_7$.  For the case of the
$CGG$ interaction we determine
\brr \d(\,-\frac{\pi}{3}\,\int C\wedge G\wedge G\,)
     &=& -\frac{\pi}{3}\,\sum_{i=1}^2\, \ft14\,(b-1)\,b^2\,\int_{M^{10}_i}\,
     \w_{2\,(i)}^1\wedge I_{4\,(i)}\wedge I_{4\,(i)} \nonumber\\
     & & -\frac{\pi}{3}\,\sum_{i=1}^{32}\,\ft12\,(b-1)\,b\,g_i\,\int_{M^6_i}\,
     \w_{2\,(i)}^1\wedge I_{4\,(i)} \,.
\label{a10}\err
To obtain this result, we note that since $G$ is gauge invariant, 
only the variation 
of the factor $C$ on the left-hand side of (\ref{a10}) contributes.
Using the explicit result (\ref{dc}), this tells us that
$\d\int CG^{\,2}=(b-1)\int_{M^{10}_i} \w_{2\,(i)}^1(G^{\,2}\,|)$, 
where the bar indicates that $G^{\,2}$ is
evaluated on $M^{10}_i$. Since 
$C_{ABC}|=0$, only the terms $\der_{[A}C_{BC](11)}$ contribute to
$dC|$, so that $dC|$ necessarily includes a $dx^{11}$ factor. 
Since both $\d C$ and $\d^{(1)}_{M^{10}_i}$ also contain $dx^{11}$ factors, 
we can therefore neglect the first two terms on the right hand side of
(\ref{gsolve}) when evaluating $G^{\,2}|$ (because $dx^{11}\wedge dx^{11}=0$). 
As a result, $G^{\,2}|$ is
proportional to $\t_{(i)}^{\,2}=1$ so that the product $G^{\,2}$ is 
well-defined on $M^{10}_i$.

It is interesting to compare equation (\ref{a10}) with the analogous
expression (\ref{cgghw}) from the case of the $S^1/{\bf Z}_2$ orbifold.
In the present case, we find the same expression for the inflow to the
two ten-planes, but we also find an additional contribution
localized on the six-planes $M^6_i$ invariant under 
the full ${\bf Z}_2\times {\bf Z}_2$ group.

Similarly, for the case of the $GX_7$ interaction we determine 
\brr \d(\,\int G\wedge X_7) 
     = -\sum_{i=1}^{2}\,\int_{M^{10}_i}\,I_{4\,(i)}\wedge X_6^1
     -\sum_{i=1}^{32}\,g_i\,\int_{M^6_i}\,X_6^1 \,.
\label{a6}\err
To obtain (\ref{a6}) 
we have integrated by parts and used the Bianchi identity (\ref{bianchi1}).  
As in the case of the $CGG$ inflow, we find the same contribution 
as found in the case of the $S^1/{\bf Z}_2$ orbifold, (\ref{gx7hw}) , 
but we also find an additional contribution localized on $M^6_i$.

The anomaly inflow can be described by a pair of twelve-forms 
$I_{12}({\rm inflow})_i$ describing the
anomaly on the ten-planes $M^{10}_i$, and by a set of 
thirty-two eight-forms $I_8({\rm inflow})_i$
describing the anomaly on the six-planes $M^6_i$. 
These are the objects which give rise to
the sum of (\ref{a10}) and (\ref{a6}) upon descent.
Thus,
\brr I_{12}({\rm inflow})_i
     &=& -\frac{\pi}{12}\,(b-1)\,b^2\,I_{4\,(i)}^{\,3}
     -I_{4\,(i)}\wedge X_8 \nonumber\\[.1in]
     I_8({\rm inflow})_i
     &=& -\frac{\pi}{6}\,(b-1)\,b\,g_i\,I_{4\,(i)}^{\,2}
     -g_i\,X_8 \,.
\label{if}\err
One might assume that these inflow terms should cancel against the
one-loop anomaly similarly to the ${\bf Z}_2$ orbifold
anomalies described in sections 3 and 4.  This turns out to
be only partially true.  Such cancellation does occur on the
fixed ten-planes, but the six-planes are more subtle.  In fact,
there is another inflow contribution, 
anticipated in section 5 which contributes to the six-dimensional 
anomaly. 

\vspace{.1in}
\noindent
{\it An ``I-brane" anomaly:}\\[.1in]
The fixed seven-planes can support the special kind of 
Chern-Simons interaction described by equation (\ref{d2cs}), 
\brr S_{CS}(M^7)=\sum_{i=1}^{16}\,\int_{M^{11}}
     \d^{\,(4)}_{M^7_i}\wedge
     G\wedge Y_{3\,(i)}^0 \,,
\label{cs1}\err
where $Y_{3\,(i)}^0$ is a Chern-Simon's three-form which can include 
Lorentz as well as a gauge pieces arising
from twisted seven-dimensional Yang-Mills matter.  Equation 
(\ref{cs1}) describes an electric coupling of $G$ to 
each of the sixteen fixed seven-planes.
If we define $Y_{4\,(i)}\equiv dY_{3\,(i)}^0$, then the most 
general $Y_{4\,(i)}$ is given by
\brr Y_{4\,(i)}=\frac{1}{(2\pi)^3\,4!}\frac{3}{32}\,\bpl\eta\,{\rm tr}\,R^2
     +\rho\,{\rm tr}\,{\cal F}_i^2\,\bpr \,,
\label{y4def}\err 
where $R$ is the eleven-dimensional curvature and 
${\cal F}_i$ is a {\it seven}-dimensional field strength
for vector fields propagating on $M^7_i$
with values in the adjoint of a group ${\cal G}_i$, to be determined.  
The parameters $\eta$ and $\rho$ are 
arbitrary rational coefficients.  The separate numerical prefactor
in (\ref{y4def}) has been chosen to simplify expressions later on.
The interaction (\ref{cs1}) allows for an anomaly contribution
of the sort described by equation (\ref{yib}).  This would arise
at a six-dimensional intersection between a ten-plane and 
a seven-plane as interplay between the ten-dimensional
magnetic coupling in the Bianchi
identity (\ref{bianchi1}) and the seven-dimensional electric  
coupling implied by (\ref{cs1}).  The anomaly contribution 
on the six-plane intersection of the $i$th ten-plane and the 
$j$th seven-plane is then given by
\brr I_8(IB)_{ij}=I_{4\,(i)}\wedge Y_{4\,(j)} \,,
\label{i8ib2}\err 
where $i$ takes either the value 1 or 2, while $j$ can assume
any value from 1 to 16.  An alternate labeling scheme involving
only a single index facilitates analysis later on.
So we adopt the convention of labeling the thirty-two six-planes
with a single index, so that (\ref{i8ib})
is rewritten as
\brr I_8(IB)_i=I_{4\,(i)}\wedge Y_{4\,(i)} \,,
\label{i8ib}\err 
where $i$ now assumes any value from 1 to 32.  In this case,
$I_{4\,(i)}$ is taken on the particular ten-plane $M^{10}_i$
which intersects $M^6_i$
while $Y_{4\,(i)}$ is taken on the particular seven-plane which 
also intersects $M^6_i$.

The combination of the inflow contributions shown in
(\ref{if}) and the ``I-brane" contribution shown in
(\ref{i8ib}) must properly conspire with the one-loop anomalies
in order that the theory be consistent.  We proceed to 
analyze first the cancellation of the ten-dimensional anomaly and then
the six-dimensional anomaly.

\subsection{The Ten-Dimensional Anomaly}
The anomaly inflow to the fixed ten-planes was computed 
above, and expressed in (\ref{if}) as $I_{12}({\rm inflow})_i$.
The two terms in $I_{12}({\rm inflow})_i$ arise due to the classical 
variation of the $CGG$ and the $GX_7$ terms respectively.
So we can write $I_{12}({\rm inflow})_i=I_{12}(CGG)_i+I_{12}(GX_7)_i$.
Another contribution
arises from one-loop diagrams.  In fact, since the ten-planes are
only invariant under the element $\a$, and since (along with the 
unit operator) $\a$ describes precisely the same ${\bf Z}_2$
group which defines the orbifold analyzed in section 3,
it follows that the computation of the one-loop anomaly in that
section applies here as well. So we do not need
to perform a separate computation of the ten-dimensional one-loop
anomaly; it is given by equation (\ref{fachw}).
Thus, the three contributions to the ten-dimensional anomaly
are given by the following polynomials,
\brr I_{12}(CGG)_i &=& -\frac{\pi}{12}\,(b-1)\,b^2\,I_{4\,(i)}^{\,3} 
     \nonumber\\[.1in]
     I_{12}(GX_7)_i &=& -I_{4\,(i)}\wedge X_8 \nonumber\\[.1in]
     I_{12}({\rm 1\,loop})_i &=& \frac{\pi}{3}\,I_{4\,(i)}^{\,3}
     +I_{4\,(i)}\wedge X_8 \,,
\label{tenp}\err
where the inflow contributions were computed above, 
and the one-loop contribution was computed in 
section 3 and given as equation (\ref{fachw}).
Not suprisingly, the inflow terms $I_{12}(CGG)_i$ and
$I_{12}(GX_7)_i$ are also the same as those derived in section 3,
given in equation (\ref{inflowhw}).

The total gravitational anomaly is given by the sum of all three
contributions in (\ref{tenp}),
$I_{12}({\rm total})_i=I_{12}({\rm 1\,loop})_i+I_{12}(CGG)_i+I_{12}(GX_7)_i$.
We require that this total anomaly vanish.  Nicely, the second term
of $I_{12}({\rm 1\,loop})_i$ is exactly canceled by $I_{12}(GX_7)_i$.
The first term of $I_{12}({\rm 1\,loop})_i$ is exactly canceled by 
$I_{12}(CGG)_i$ provided $b$
satisfies the cubic equation $b^3-b^2-4=0$.  This equation has one real root,
so anomaly cancellation uniquely selects 
\brr b=2 \,.
\err  
This value of $b$ is fixed by the consistency requirements. 
It is gratifying that this requirement is satisfied by a 
rational (indeed, integer) value for $b$. 
The sub-analysis of the ten-dimensional anomaly 
in the ${\bf Z}_2\times {\bf Z}_2$ case is precisely the same as that 
given in section 3 for the simpler ${\bf Z}_2$ orbifold. 

\subsection{The Six-Dimensional Anomaly}
We are mostly concerned with the additional anomaly localized
on the six-dimensional fixed-planes $M^6_i$ invariant
under the element $\b$.  Since (along with the unit element)
$\b$ describes precisely the same ${\bf Z}_2$
group which defines the orbifold analyzed in section 4, one might 
believe that we can rely on the computation in section 4
in the same manner that we relied on the computation of section 3
to obtain the ten-dimensional anomaly.  But things are not so simple
in this case because $M^6_i$ are not just 
invariant under $\b$ but are, in fact, invariant under $\a$
and $\a\b$ as well!  So the six-planes are invariant under the
entire ${\bf Z}_2\times {\bf Z}_2$ group.  As a result, there are
crucial differences from the ${\bf Z}_2$ orbifold analyzed in section
4.  For one thing, the additional ${\bf Z}_2$ projects out yet another half
of the bulk supercharges, so that the local supersymmetry on
these planes is D=6 N=1 rather than D=6 N=2 as in section 4.
This not only changes the untwisted spectrum, and therefore the one-loop
anomaly from that of the ${\bf Z}_2$ case, but it also means
that we have the freedom to include a richer twisted spectrum of
additional local states due to the richer structure of N=1
supersymmetry. 

\vspace{.1in}
\noindent
{\it Anomaly inflow:}\\[.1in]
Anomaly inflow to the fixed six-planes was computed 
above, and expressed as $I_{8}({\rm inflow})_i$ in (\ref{if}),
where the two terms arise due to the classical 
variation of the $CGG$ and the $GX_7$ terms respectively.
So we can write $I_{8}({\rm inflow})_i=I_{8}(CGG)_i+I_{8}(GX_7)_i$.
Another contribution $I_8(IB)_i$ arises from the ``I-brane" mechanism 
also described above and given as equation (\ref{i8ib}).
We set $b=2$ as required by the removal of the anomaly on the
ten-planes. Thus, the three inflow contributions to the 
six-dimensional anomaly are given by the following polynomials,
\brr {I}_8(GX_7)_i &=&  -g_i\,X_8
      \nonumber\\[.1in]
     {I}_8(CGG)_i &=&  -\ft13\,\pi\,I_{4\,(i)}^{\,2}
     \nonumber\\[.1in]
     {I}_8(IB)_i &=& I_{4\,(i)}\wedge Y_{4\,(i)}  \,,
\label{inflow6}\err
where $X_8$ is given in (\ref{x8def}), while $I_{4\,(i)}$
is given in (\ref{i4def}) and $Y_{4\,(i)}$ is given in
(\ref{y4def}).  As explained earlier, the notation is such that
$i$ assumes any value from 1 to 32.

The combination of the three contributions shown in
(\ref{inflow6}) must conspire with the one-loop anomalies
in order that the theory be consistent.  We proceed to 
analyze the one-loop contributions, and then discuss the 
conspiracy which renders the theory consistent.

\vspace{.1in}
\noindent
{\it The one-loop anomaly:}\\[.1in]
The quantum contributions consist of three separate pieces:

A contribution $I_8(SG)_i$ arises from the bulk supergravity 
due to the coupling of chiral projections 
of the gravitino to $SO(5,1)$ currents associated with
diffeomorphisms of the fixed six-planes.  On a 
given six-plane, only the components
$C_{(11)\mu\nu}$ and $C_{(11)ij}$ (where $i,j=1,...,4$)
survive the ${\bf Z}_2\times {\bf Z}_2$ projection from the
three-form $C_{IJK}$. These contribute
one two-form and six scalars, while the eleven-dimensional
metric supplies a six-dimensional metric and
eleven more scalars corresponding to $g_{(11)(11)}$
and $g_{ij}$.  So the bosonic untwisted spectrum consists of
a metric tensor, one two-form and seventeen scalars.  These
organize along with the surviving fermions into
a D=6 N=1 supergravity multiplet coupled to four N=1 hypermultiplets
and one N=1 tensor multiplet.  

Since the total anomaly is distributed equally over the 
thirty-two six-planes,
and since we can only apply the index theorem results in the 
(small-radius) limit when all six-planes coincide, we conclude that
the anomaly on a given hyperplane is 1/32 of that described by the
index theorem results using the untwisted spectrum associated with the
bulk fermions.
Collectively, these involve
one chiral spin 3/2 field, five antichiral spin 1/2 fields 
and one each of self-dual and anti-self-dual tensors.  
The anomalies due to the tensors cancel each other, so that 
\brr I_8(SG)_i=\ft{1}{32}\,\bpl\,I_{GRAV}^{(3/2)}\,(R)
     -5\,I_{GRAV}^{(1/2)}\,(R)\,\bpr \,.
\label{sgterm}\err

Another contribution, $I_8(E_8)_i$, arises from the $E_8$ matter propagating
on the fixed ten-planes.
To begin with, we assume that $\b$ acts trivially on the ten-dimensional
vector multiplets, so that the $E_8$  gauge group is not broken by 
the orbifold action.
A given ten-dimensional $E_8$ vector supermultiplet decomposes into 
a D=6 N=2 $E_8$ vector multiplet, which further decomposes into
an N=1 vector multiplet and an N=1 hypermultiplet.  The first of these 
involves chiral gauginos while the second involves antichiral
hyperinos.  The anomalies due to these two factors would cancel 
against each other.  However, 
the six-planes are fixed under {\it both} ${\bf Z}_2$ factors $\a$ and $\b$.  
The second ${\bf Z}_2$, denoted $\b$, acts on the $E_8$ supermatter to project
out the N=1 hypermultiplet, leaving a contribution only from the 
N=1 $E_8$ vector multiplet, which is anomalous.  Since there are sixteen
fixed six-planes within a given ten-plane,
the contribution $I_8(E_8)_i$ localized on a given six-plane is 1/16 of 
that described by the index theorem results pertaining to
(ten-dimensional) chiral $E_8$ gauginos, so that
\brr I_8(E_8)_i=\ft{1}{16}\,\bpl\,248\,I_{GRAV}^{(1/2)}(R)
     +I_{MIXED}^{(1/2)}\,(R,F_i)_{ADJ}
     +I_{GAUGE}^{(1/2)}\,(F_i)_{ADJ}\,\bpr \,,
\label{e8term}\err
where $F_i$ takes values in the adjoint {\bf 248} representation of 
the $E_8$.

The element $\b$ can also act nontrivially on the $E_8$ vectors, breaking
the group to a maximal subgroup.  For instance,  the {\bf 248} decomposes
into $E_7\times SU(2)$ representations 
as $({\bf 133},1)\oplus (1,{\bf 3}) \oplus ({\bf 56},{\bf 2})$.
We could realize the ${\bf Z}_2$ on the $E_8$ fields 
to project out the six-dimensional hypermultiplets from the
$({\bf 133},1)\oplus (1,{\bf 3})$ fields and project out the six-dimensional
vector multiplets from the $({\bf 56},{\bf 2})$ fields.  In this case,
we would be left with $E_7\times SU(2)$ adjoint vectors and 
112 hypermultiplets transforming as $({\bf 56},{\bf 2})$.  
Another possibility would break $E_8$ to $Spin(16)$,
leaving us with the adjoint {\bf 120} coupled to a hypermultiplet
in the {\bf 128} spinor representation.  The two possibilities
described by the $E_7\times SU(2)$ and $Spin(16)$ cases correspond to the 
{\it only} ${\bf Z}_2$ subgroups of $E_8$ \cite{sixa}.
Therefore there are only three
possibilities, $E_8\to E_8$, $E_8\to E_7\times SU(2)$ and 
$E_8\to Spin(16)$.  We will analyze only the first possibility in 
detail, and make comments about the other two afterwards. 

The third contribution $I_8({\cal G}_i)$
arises from twisted matter which we are free to add to the fixed planes.
In fact, we are free to add twisted matter of two significantly
different sorts.  On the one hand, we can include {\it six}-dimensional
fields propagating on any or all of the thirty-two
six-planes, consisting of some number of
D=6 N=1 vector, hyper and/or tensor multiplets.  On the other hand, we
are also free to add {\it seven}-dimensional fields propagating
on any or all of the sixteen seven-planes, consisting of
seven-dimensional vector multiplets.  
Each of these possibilities will contribute to the 
six-dimensional anomaly.  In the first case,
the chiral fields living in the six-dimensional multiplets
will couple anomalously to the six-dimensional Lorentz and gauge currents.
In the second case, even though the seven-dimensional gauginos will
not contribute to an anomaly on the seven-planes (since they are non-chiral),
they {\it will} couple anomalously to six-dimensional currents on the 
subplanes fixed by the entire ${\bf Z}_2\times {\bf Z}_2$ group.
These are, of course, the same six-planes where six-dimensional 
twisted fields can propagate, represented by the solid dots in figure 1.

The reason why the seven-dimensional fields can contribute to the
six-dimensional anomaly mirrors the way in which
{\it ten}-dimensional $E_8$ gauginos contribute to the six-dimensional
anomaly.  As described above, each ten-dimensional chiral gaugino decomposes
into one six-dimensional chiral gaugino and one six-dimensional
antichiral hyperino.  In that case, the extra ${\bf Z}_2$ projection
which leaves the six-planes fixed serves to remove the hyperinos, so that
there is a net (six-dimensional) chirality to the projected 
fermions.  Similarly,
a seven-dimensional gaugino also decomposes into a six-dimensional
gaugino and a six-dimensional hyperino \footnote{This is easy to see from the
bosonic components of the seven-dimensional Yang-Mills multiplet, which
comprises three scalars and one vector; upon torroidal compactification of
one dimension, the seven-dimensional vector will give a six-dimensional
vector and a fourth scalar, giving the bosonic components of 
one D=6 N=1 hyper and one D=6 N=1 vector multiplet}.  Once again
the extra ${\bf Z}_2$ projection will remove the hyperinos,
so that the projected fields have a net (six-dimensional) chirality.

We first consider the case where we add {\it six}-dimensional twisted matter.
If, on the $i$th six-plane we add $n_{Vi}$ vector multiplets,
$n_{Hi}$ hypermultiplets,
and $n_{Ti}$ gauge-singlet tensor multiplets, the relevant anomaly
is given by
\brr I_8({\cal G}_i) &=& (n_V-n_H-n_T)_i\,I_{GRAV}^{(1/2)}(R)
     -n_{Ti}\,I_{GRAV}^{({\rm 3-form})}(R) \nonumber\\[.1in]
     & & +I_{MIXED}^{(1/2)}(R,{\cal F}_i)_{ADJ,R}
     +I_{GAUGE}^{(1/2)}({\cal F}_i)_{ADJ,R} \,,
\label{i86}\err
where the mixed and pure gauge anomalies involve field strength tensors
${\cal F}_i$ taking values in both the adjoint (in the case of vector 
multiplets), and in the $R$ representation (for the case of hypermultiplets).
To make sense of equation (\ref{i86}), we should use the polynomials
given in (\ref{grav6}) and (\ref{rf6}), and replace the 
${\rm trace}\,{\cal F}_i^2$
contribution in the mixed anomaly and the 
${\rm trace}\,{\cal F}_i^4$ contribution
in the gauge anomaly as
\brr {\rm trace}\,{\cal F}_i^n\equiv {\rm Tr}\,{\cal F}_i^n
     -\sum_\a h_\a\,{\rm tr}_\a\,{\cal F}_i^n \,,
\label{tracer}\err
where ${\rm Tr}$ is an adjoint trace, $h_\a$ is the number 
of hypermultiplets transforming in the $R_\a$ representation, and 
${\rm tr}_\a$ is a trace over the $R_\a$ representation.  
Note that the total number of vector multiplets is 
$n_{Vi}={\rm dim}\,({\cal G}_i)$
while the total number of hypermultiplets is
$n_{Hi}=\sum_\a\,h_\a\times{\rm dim}\,(R_\a)$.
The relative minus sign in (\ref{tracer}) reflects the 
{\it anti}chirality of the hyperinos.

Now consider the case where we add {\it seven}-dimensional twisted
matter.  If on a given $\b$-invariant seven-plane $M^7_i$ 
we add $\tilde{n}_{Vi}$ vector multiplets in the adjoint of 
${\cal G}_i$,
these will contribute to anomalies on the embedded
$\a$- and $\b$-invariant six-planes due to the extra $\a$
projection.  In the simplest case, $\a$ will 
remove all but a six-dimensional hypermultiplet from the 
seven-dimensional fields.  There are other possibilities where 
$\a$ breaks ${\cal G}_i$ to a maximal subgroup analgous to the 
situation involving the $\b$ projection on the $E_8$ fields discussed
above.  Once again, we will consider first the simplest case, 
where $\a$ does not break ${\cal G}_i$ and comment on the other 
possibilities later.  

Since there are {\it two}
fixed six-planes within a given seven-plane, it follows by reasoning
described above that the contribution to the six-dimensional anomaly
on each of the two six-planes due to the seven-dimensional gauginos
will be given by
\brr I_8({\cal G}_i) &=& \frac{1}{2}\,\bpl\,
     \tilde{n}_{Vi}\,I_{GRAV}^{(1/2)}(R)
     +I_{MIXED}^{(1/2)}(R,{\cal F}_i)_{ADJ}
     +I_{GAUGE}^{(1/2)}({\cal F}_i)_{ADJ}\,\bpr \,,
\label{i87}\err
where the factor of 1/2 arises because the anomaly is
equally distributed over the two fixed six-planes.  
To make
sense of equation (\ref{i87}), we should use the explicit polynomials
given in (\ref{grav6}) and (\ref{rf6}), making the substitution 
indicated in (\ref{tracer}).
If ${\cal G}_i$ is unbroken by $\a$, then only the adjoint vectors
will contribute, so the $h_\a$ would be zero.
If ${\cal G}_i$ is broken to a maximal subgroup 
by $\a$, then vectors in the adjoint of the subgroup will contribute
along with some number of hypermultiplets.  

It is also possible to include seven-dimensional 
adjoint ${\cal G}_i$ vectors
and to include additional {\it six}-dimensional hypermultiplets which 
also transform under ${\cal G}_i$.  In this case,  the anomaly due to the
vectors would be given by (\ref{i87}), while that due to the hypers would
be given by (\ref{i86}).  

In all cases, the only distinguishing qualification is a division 
by two, as seen in equation (\ref{i87}), for the anomaly
due to any seven-dimensional field.  Any purely six-dimensional field 
contributes to the anomaly without such a division, as in 
(\ref{i86}).  It turns out that further considerations 
concerning the factorizability
of the anomaly polynomial {\it require} factors of two in such a way
that implies the existence of
seven-dimensional twisted matter.  This is one of the
essential points of this paper.

Our strategy is to include unspecified twisted states in a
sufficiently powerful way that
anomaly cancellation will select both gauge groups and also
the {\it dimensionality} of the appropriate fields for us.  
To do this, we need a certain economy which is had by encorporating the
various possibilities involving six- and seven-dimensional states
in a unified package.  This is facilitated in an obvious way by
writing the complete twisted anomaly precisely as in
equation (\ref{i86}), but with two important distinctions. The first
distinction is that, since seven-dimensional fields 
contribute one-half the anomaly on a given six-plane as six-dimensional
fields, we consider $n_{Vi}$ in (\ref{i86}) to include this
potential divisor.  Thus, $n_{Vi}\equiv{\rm dim}({\cal G}_i)/\mu$ where 
$\mu$ is 1 or 2 depending on the dimensionality of the fields in question,
\brr \mu\,=\,\left\{\begin{array}{lcl} 
     \,1 & \hspace{.1in} \longleftarrow &
     \hspace{.2in} {\rm six\!-\!dimensional\,\,fields} \\
     \,2 & \hspace{.1in} \longleftarrow &
     \hspace{.05in} {\rm seven\!-\!dimensional\,\,fields} 
     \end{array}\right. \,.
\label{mudef}\err

The second distinction is that when we substitute the 
anomaly polynomials (\ref{grav6}) and (\ref{rf6}) we should replace
the ${\rm trace}\,{\cal F}_i^2$ and ${\rm trace}\,{\cal F}_i^4$ 
contributions not
with (\ref{tracer}) but rather with the obvious extension
\brr {\rm trace}\,{\cal F}_i^n\,\equiv\,\frac{1}{\mu}\,{\rm Tr}\,{\cal F}_i^n
     -\sum_\a h_\a\,{\rm tr}_\a\,{\cal F}_i^n \,,
\label{trace2}\err
where ${\rm Tr}$ is an adjoint trace, $h_\a$ is the number 
of six-dimensional hypermultiplets transforming in the 
$R_\a$ representation, and 
${\rm tr}_\a$ is a trace over the $R_\a$ representation. In this
generalized formulation, the fact that seven-dimensional fields 
contribute {\it one-half} of the mixed and gauge anomalies 
as do six-dimensional fields is encorporated in the parameter $\mu$.  

The complete quantum anomaly on a given six-plane is given by the 
sum of (\ref{sgterm}), (\ref{e8term}) and (\ref{i86}).  To ease our
analysis we repeat the three quantum contributions here,
\brr I_8(SG)_i &=& \ft{1}{32}\,\bpl\,I_{GRAV}^{(3/2)}\,(R)
     -5\,I_{GRAV}^{(1/2)}\,(R)\,\bpr  
     \nonumber\\[.1in]
     I_8(E_8)_i &=& \ft{1}{16}\,\bpl\,248\,I_{GRAV}^{(1/2)}(R)
     +I_{MIXED}^{(1/2)}\,(R,F_i)_{ADJ}
     +I_{GAUGE}^{(1/2)}\,(F_i)_{ADJ}\,\bpr 
     \nonumber\\[.1in]
     I_8({\cal G}_i) &=& (n_V-n_H-n_T)\,I_{GRAV}^{(1/2)}(R)
     -n_T\,I_{GRAV}^{({\rm 3-form})}(R) \nonumber\\[.1in]
     & & +I_{MIXED}^{(1/2)}(R,{\cal F}_i)_{ADJ,R}
     +I_{GAUGE}^{(1/2)}({\cal F}_i)_{ADJ,R} \,.
\label{quantum}\err
The $I_8(SG)_i$ contribution arises from chiral projections of the
bulk supergravity fields; the division by 32 reflects the fact that
this anomaly is equally distributed over the thirty-two $M^6_i$. 
The $I_8(E_8)_i$ contribution arises from chiral
projections of the ten-dimensional $E_8$ gauginos; in this case the
division by 16 reflects the fact that each ten-plane has sixteen
embedded fixed six-planes over which this anomaly is equally distributed.
Finally,  the $I_8({\cal G}_i)$ contribution arises from the 
(yet-undetermined) twisted fields which can propagate {\it either} 
on the six-planes
or on the seven-planes, the distinction being encoded in the 
parameter $\mu$ as described above. The parameter $\mu$
is an important algebraic tool for ascertaining the dimensionality
of twisted states indicated by anomaly cancellation.

Henceforth, since it is clear that we are working on a particular
six-plane $M^6_i$ we will suppress the $i$ index on objects
like $g_i$ and $n_{H\,i}$.

\vspace{.1in}
\noindent
{\it The total anomaly:}\\[.1in]
The total anomaly on a given six-plane is given by the sum of the
three inflow contributions given in (\ref{inflow6}) and the
three quantum contributions presented in (\ref{quantum}).  Using the
anomaly polynomials given in (\ref{grav6}) and (\ref{rf6}), we can
work out the full anomaly polynomial.  This will have many terms.
However, only the term proportional to ${\rm tr}\,R^4$ will
not be factorizable.  This term has a coefficient proportional
to $(n_H-n_V+29n_T-30g-23)$.
Because the ${\rm tr}\,R^4$ term cannot factorize, it cannot
be removed by a Green-Schwarz mechanism using a local tensor
field and therefore must vanish identically.  This poses an important
constraint on the allowed twisted states, given by  
\brr n_H-n_V=30g+23-29n_T \,.
\label{gcon}\err
If we sum this equation over the thirty-two six-dimensional fixed-planes,
impose the constraint (\ref{cons2}),
and use the following expressions for the total number of
hyper, vector and tensor multiplets,
\brr N_H &=& 4+N_5+\sum n_H \nonumber\\[.1in]
     N_V &=& 496+\sum n_V \nonumber\\[.1in]
     N_T &=& 1+N_5+\sum n_T \,,
\label{total}\err
we arrive at the ``collective" constraint
$N_H-N_V+29N_T=273$, which is a more familiar anomaly requirement.
The numbers 4, 496 and 1 which appear in (\ref{total})
describe the untwisted contributions,
since there are 4 untwisted hypers and one untwisted tensor coming from the 
bulk supergravity and there are 496 untwisted vectors coming from the
two $E_8$ factors.  Also,  each fivebrane contributes one hypermultiplet
and one tensor multiplet.

In the anomaly polynomial, terms corresponding to 
mixed and gauge anomalies include traces over powers of
the field strength ${\cal F}$ in precisely the combinations shown in
equation (\ref{trace2}).  We should represent all traces
in terms a fundamental representation.
There are identities which relate traces over any representation
of a given simple gauge factor in terms of traces over other 
representations.  The dimensions of the most useful
representations for the simple gauge groups is given in table 2, and
a useful tabulation of the trace relations is
given in table 3 for the non-exceptional groups and in table 4
for the exceptional groups. 

The anomaly polynomial contains a term proportional to 
${\rm tr}\,{\cal F}^4$, which
must factorize as $({\rm tr}\,{\cal F}^2)^2$.  This is done automatically
for the exceptional groups and for $SU(2)$ and $SU(3)$, since these
groups do not have an independent fourth-order Casimir operator.
But for other simple groups this factorization requires a 
conspiracy involving the multiplicities of the hypermultiplet
representations.  If we assume that the proper
factorization of the ${\rm tr}\,{\cal F}^4$ term occurs, then in
all cases the mixed and gauge anomalies will involve the following
two factors
\brr \frac{1}{6}\,\bpl\,\frac{1}{\mu}\,{\rm Tr}\,{\cal F}^2
     -\sum_\a h_\a\,{\rm tr}\,{\cal F}^2 \,\bpr
     &\equiv& -X\,{\rm tr}\,{\cal F}^2
     \nonumber\\[.1in]
     \frac{2}{3}\,\bpl\,\frac{1}{\mu}\,{\rm Tr}{\cal F}^4
     -\sum_\a h_\a\,{\rm tr}{\cal F}^4 \,\bpr
     &\equiv& -\,Y\,(\,{\rm tr}\,{\cal F}\,)^2 \,,
\label{xydef}\err
where we have {\it defined} $X$ and $Y$ as the generic coefficients
of the $({\rm tr}\,{\cal F}^2)^2$ term (which multiplies ${\rm tr}\,R^2$
in the anomaly polynomial to form a mixed anomaly) 
and the ${\rm tr}\,{\cal F}^4$ term (which constitutes the pure
gauge anomaly).  It is possible to determine the coefficients $X$ and $Y$
for each possible simple gauge factor by applying the trace relations given in 
tables 3 and 4 to the expressions in (\ref{xydef}).
The numbers $X$ and $Y$ for generic choice of simple gauge factor and
arbitrary hypermultiplet representation are given in table 5.

We now substitute (\ref{gcon}) into the anomaly polynomial
to remove the ${\rm tr}\,R^4$ term.  We also use (\ref{gcon}) to
replace the combination $n_V-n_H$ in terms of $n_T$ and $g_i$.
It is then a matter of straightforward algebra to determine the 
final form of the anomaly polynomial.  
For convenience, we define a hatted polynomial by 
removing a common prefactor,
\brr I_8({\rm total})\equiv
     \frac{1}{(2\pi)^3\,4!}\,\frac{3}{2}\,\hat{I}_8({\rm total}) \,.
\err
Now if we compute the total anomaly by summing up the three inflow
contributions (\ref{inflow6}) and the three quantum contributions
(\ref{quantum}), then remove the ${\rm tr}\,R^4$ term by imposing
(\ref{gcon}), and finally rewrite all traces over twisted gauge
factors using (\ref{xydef}) we detemine the complete anomaly to
be given by the following expression,
\brr \hat{I}_8(\,{\rm total}\,) &=&
     \ft{1}{32}(1+\eta+\ft{8}{3}g-4n_T)\,(\,{\rm tr}\,R^2\,)^2
     \nonumber\\[.1in]
     & & +\ft{1}{16}(5-\eta+\ft83 g)\,{\rm tr}\,R^2\wedge{\rm tr}\,F^2
     -\ft{1}{24}(9+4g)\,(\,{\rm tr}\,F^2\,)^2
     \nonumber\\[.1in]
     & & -\ft{1}{16}\rho\,{\rm tr}\,F^2\wedge{\rm tr}\,{\cal F}^2
     +(\ft{1}{32}\rho-X)\,{\rm tr}\,R^2\wedge{\rm tr}\,{\cal F}^2
     +Y\,(\,{\rm tr}{\cal F}^2\,)^2 \,,
\label{theanomaly}\err
where $F$ is the $E_8$ gauge field and ${\cal F}$ is the twisted
gauge field.  The expression (\ref{theanomaly}) includes
unspecified local magnetic charge $g$, an arbitrary number of
local tensor fields $n_T$, an arbitrary ``I-brane" contribution 
parameterized by $(\eta,\rho)$, and an arbitrary twisted spectrum
parameterized by $(X,Y)$.

The anomaly (\ref{theanomaly}) cannot vanish identically.  This can
be seen by the following simple observations.  Cancelation of the 
$({\rm tr}\,F^2)^2$ term would require $g=-9/4$.  Using this result,
cancellation of the ${\rm tr R}^2\wedge{\rm tr}F^2$ term would
require $\eta=-1$.  Cancellation of the $({\rm tr}\,R^2)^2$ term would
then require that $n_T=-3/2$, which is not positive,
and therefore not realizable.  So we must resort to a local
Green-Schwarz mechanism to cancel the anomaly.  
In the following subsection we attempt to implement this.

\subsection{Factorizing the anomaly}
Since the anomaly cannot vanish identically, it can only be removed via
a Green-Schwarz mechanism realized locally (ie: on the particular 
six-plane $M^6_i$) through the local coupling of at least one tensor field.   
Since the fixed six-planes have $N=1$ supersymmetry, any such
tensor would have an anti self-dual field strength.  
Therefore, the mechanism necessitates that
the anomaly polynomial (\ref{theanomaly}) factorize into a sum 
of perfect squares
\footnote{We thank to Stephan Theisen for discussions on this point.}, 
one term for each tensor field.  The minimal case would
involve one twisted tensor field.  In that case, we would impose that
(\ref{theanomaly}) factorizes as
\brr \hat{I}_8({\rm total}) =
     \ft{1}{32}\,r\,(\,{\rm tr}\,R^2
     -u\,{\rm tr}\,F^2-t\,{\rm tr}\,{\cal F}^2\,)^2 \,,
\label{aform}\err
where $r$, $u$ and $t$ are rational coefficients to be determined.
We discuss the possibility of more than one tensor field below.
The one case not covered by the parameterization (\ref{aform}) is
where the $({\rm tr}\,R^2)^2$ term vanishes but other terms
do not.  This possibility is considered separately and found
not to be relevant.  The reason why we include $r$ as an overall
coefficient, rather than inside the brackets multiplying the
${\rm tr}\,R^2$ term, is because the analysis is more tractable
this way.
Equating equation (\ref{aform}) with (\ref{theanomaly}) generates 
six relations encapsulating the factorization requirement, 
\brr r &=& 1+\eta+\ft83\,g-4\,n_T \hspace{.35in} 
     r\,t = 16\,X-\ft12\,\rho
     \nonumber\\[.1in]
     r\,u &=& -5+\eta-\ft83\,g \hspace{.6in}
     r\,u\,t=-\rho
     \nonumber\\[.1in]
     r\,u^2 &=& -12-\ft{16}{3}\,g \hspace{.8in}
     r\,t^2=32\,Y \,.
\label{factor8}\err
Three of these (those in the left-hand column) relate to terms which 
include ${\rm tr}\,F^2$ and, therefore, concern the $E_8$ anomaly.
The other three (those in the right-hand column) relate to terms which 
include ${\rm tr}\,{\cal F}^2$ and, therefore, concern the 
twisted anomaly.  

The goal is to find rational
values for the magnetic charge $g$, the ``electric" parameters
$\eta$ and $\rho$ \footnote{The parameters $\eta$ and $\rho$ have an
``electric" nature since they determine the electric couplings described
by (\ref{cs1}) and (\ref{y4def}).} and the group-dependent parameters
$X$ and $Y$ (which are defined in (\ref{xydef})) in such a way that
(\ref{factor8}) determines rational values for $r$, $u$ and $t$.

The parameters $X$ and $Y$ have a very restricted and group-specialized
dependence on the multiplicities of the hypermultiplets.  These relationships
are exhibited in table 5.  Furthermore, $X$ and $Y$ also depend on $\mu$, the
parameter defined in (\ref{mudef}), which tells us the {\it dimensionality}
of the hyperplane on which the
twisted states live.  Therefore, even if we manage to find a solution
to ({\ref{factor8}), this is not enough, since the values of $X$ and $Y$
must be realizable for some choice of simple gauge factor, hypermultiplet
multiplicity and value $1$ or $2$ for the parameter $\mu$.  The reason
why we can specialize to simple groups is explained in the following paragraph.
``Realizable", in this instance, means that there is a choice of
nonnegative integer multiplicity $h_\a$ for hypermultiplets transforming
in some representation $R_\a$ for some choice of simple gauge 
factor ${\cal G}$.  This choice must describe precisely
the given values of both $X$ and $Y$ using the relationships listed in
table 5, for either of the cases $\mu=1$ or $\mu=2$.
There is a systematic approach that allows one to 
sift sequentially through all possibilities.

Even if the ultimate solution involves a semisimple group,
there are necessary constraints pertaining to each simple factor
of that group which coincide with those which we are analyzing.
Therefore, we do not lose generality by specializing to the
case of simple gauge groups.  Nevertheless, once we find a solution
to the factorization problem for a given simple factor, we should
explore the possibility that this group can be extended to include
additional simple factors. Other constraints discussed below may
{\it require} such an extension in generic cases.  

If one can find a solution to (\ref{factor8}) which includes
``realizable" values for $X$ and $Y$, this is still not enough. 
There remains the extra constraint (\ref{gcon}) related to the
vanishing of the irreducible part of the anomaly.  So, once we have
determined a set of rational parameters which satisfy (\ref{factor8})
{\it and} a gauge group ${\cal G}$ {\it and} hypermultiplet multiplicity 
$h_\a$ which give rise to the given $X$ and $Y$, this will determine 
the total number of twisted vector multiplets $n_V$ and the
total number of twisted hypermultiplets $n_H$.  But these
must satisfy (\ref{gcon}).  Finding a solution which meets all of 
these criteria is nontrivial.

Furthermore, there is yet another constraint.  This is the global constaint
(\ref{cons2}) which tells us that the sum of all 32 magnetic charges
plus the number of fivebranes must vanish.  This requires that 
at least one of the six-planes has nonpositive magnetic charge.  
It is, therefore, essential that we find a solution with nonpositive $g$.

\vspace{.1in}
\noindent
{\it A ``solution":}\\[.1in]
It turns out that there
is a unique solution to all of the above constraints,
for the case of nonpositive g.  
We emphasize that the equations which we have developed 
are sensitive to each of a large number of coefficients 
involved in a lengthy multifaceted analysis.  
If any one of these coefficients would change slightly,
there would be no solution at all.  
Given the complexity of the analysis, which involves a delicate
array of effects, we find the existence of one and only one solution 
to all of these constraints to be significant.  

Nevertheless there remains a paradox associated with the solution.
Specifically, the solution requires $n_T=1/2$, which is not
an integer value.  However, the mere existence of such a ``formal"
solution is remarkable for the reasons expressed above.
Furthermore, there are significant insights which derive
from this solution, and indications that the paradox itself
may not be inscrutible.  In fact, a similar paradox
attended the discoveries outlined in section 4 in the
context of the $T^5/{\bf Z}_2$ orbifold.  In that case, a 
``half-tensor" multiplet seemed to be required by anomaly cancellation 
before a missing inflow mechanism involving the magnetic charge $g_i$
was included. So, despite the described shortcoming,
we will explain the solution in more detail and speculate on
possible resolutions to the above paradox in due course.

There is another notable aspect to the solution.  
This is that the necessary magnetic charge assignment
for the six-plane $M^6_i$ in question turns out to be $g_i=-3/4$.
As discussed earlier, there are reasons to expect such a quarter-integer
value for $g_i$ in the context of this orbifold because the construction
represents a singularly deformed limit of $K3\times S^1/{\bf Z_2}$.
One can then interpret part of the local magnetic charge
as being the a remnant of the Euler number of the smooth $K3$ manifold.
In fact, this ``Euler" contribution was earlier shown to
give precisely the value $g_i=-3/4$.  

The one valid solution with nonpositive $g$ to the factorizarion
problem described by the six conditions (\ref{factor8}) 
is given by the following rational assignments for
the parameters of the problem, 
\brr (\,n_T\,,\,g\,) &=& (\,1/2\,,\,-3/4\,) 
     \nonumber\\[.1in]
     (\,\eta\,,\,\rho\,) &=& (\,-5\,,\,16\,)
     \nonumber\\[.1in]
     (\,X\,,\,Y\,) &=& (\,-1/2\,,\,-1\,)
     \nonumber\\[.1in] 
     (\,r\,,\,u\,,\,t\,) &=&  (\,-8\,,\,1\,,\,2\,) \,.
\label{solution}\err
As described above, this solution is ``validated" only if we can
reproduce the indicated values for $X$ and $Y$ using
the relationships in table 5 for some choice of
multiplicities $h_\a$ for hypermultiplets transforming
in some representation $R_\a$ for some choice of simple gauge 
factor ${\cal G}$ for either of the cases $\mu=1$ or $\mu=2$.  We must
{\it also} to do this in such a way that (\ref{gcon}) is
satisfied.  When we substitute the indicated values $g=-3/4$ and $n_T=1/2$ 
into (\ref{gcon}), we find that this extra constraint becomes
\brr n_H-n_V=-14 \,.
\label{g8}\err
So we must find a twisted gauge group which satisfies
both $(X,Y)=(-1/2,-1)$ and also (\ref{g8}).

By carefully sifting through the possibilities in table 5, one finds
a unique simple factor which solves this problem.  
This is to put twisted $SO(8)$ gauge matter
on the {\it seven}-dimensional plane $M^7_i$ (so that $\mu=2$) and to 
set the number of hypermultiplets to zero (so that $h=h_S=0$ in table 5).
In that case, we would have $n_H=0$ while $n_V={\rm dim}{\cal G}/\mu=14$, 
the division by $\mu=2$ indicating that the $SO(8)$
gauge fields are seven-dimensional, as explained above.

It is now evident how the anomaly analysis has required seven-dimensional
twisted matter despite the fact that there is no seven-dimensional
anomaly as such. It is also clear that that this kind
of mechanism generalizes to other orbifolds involving
intersecting orbifold planes. 

Given the values of $r$, $u$ and $t$ which we have found
in (\ref{solution}), we can write down the form of the anomaly.
Substituting these values into (\ref{aform}) we determine
\brr \hat{I}_8({\rm total}) =
     -\ft{1}{4}\,(\,{\rm tr}\,R^2
     -\,{\rm tr}\,F^2-2\,{\rm tr}\,{\cal F}^2\,)^2 \,.
\label{aaaa}\err
This is the anomaly which should be canceled using a 
Green-Schwarz mechanism mediated by a local tensor field.
If our solution had had $n_T=1$, it would be clear how to
realize this; the three-form field strength would satisfy the
Bianchi identity $dH\propto{\rm tr}\,R^2-{\rm tr}\,F^2-2{\rm tr}\,{\cal F}^2$,
and the tensor dynamics would include a Chern-Simon's interaction
proportional to
$\int_{M^6_i}B\wedge({\rm tr}\,R^2-{\rm tr}\,F^2-2{\rm tr}\,{\cal F}^2)$\,.
The self-duality of $H$ would be maintained because the magnetic
and electric couplings implied by these are the same.  
Due to the modified $dH$ Bianchi identity, the tensor $B$ would transform 
in just such a way that the resulting transformation of the Chern-Simon's
interaction would cancel the anomaly (\ref{aaaa}).  However, since
our ``formal" solution has $n_T=1/2$, it is unclear whether we actually
have a tensor whose dynamics could 
include these modifications.

The global constraint (\ref{cons2}) is satisfied for our ``solution"
by assuming that each of the 32 fixed points exhibits identical
behavior.  Thus, we would have a charge of $g_i=-3/4$ for each of 
the thirty-two six-planes $M^6_i$, impling $\sum_i\,g_i=-24$.
Equation (\ref{cons2}) would then be balanced by including 
24 independent fivebranes ``wandering" in the bulk.  We remark that
in this case the magnetic charge associated with the fixed-planes
is identical to that ascribable to the ``pinched" curvature of the
singularly-deformed $K3$ manifold, as explained early in this section.

\vspace{.1in}
\noindent
{\it Additional tensors:}\\[.1in]
In principle, we can involve more than one twisted tensor field.  
As explained above, this would imply
a weaker factorization constraint than the one indicated by
(\ref{factor8}).  With more tensor fields, we would generalize
that constraint to impose that $\hat{I}_8({\rm total})_i$ factorize
as a sum of perfect squares, one for each tensor field.  But the
possibility of reasonable solutions for the case $n_T>1$ is hampered
by the independent constraint (\ref{gcon}) necessary to remove the 
irreducible part of the local anomaly.  For the case $n_T=1$ this
relation becomes $n_H-n_V=30g-6$ which implies twisted matter 
living in relatively small gauge groups for reasonable (ie: relatively
small) values of the magnetic charge because, in that 
case, $30g-6$ is not a large number.  For example, we already described one
(formal) solution with $n_T=1/2$ and seven-dimensional twisted
gauge group ${\cal G}=SO(8)$.  With an arbitrary number of 
twisted tensors, the constraint (\ref{gcon}) becomes
$n_H-n_V=30g+23-29n_T$, which quickly becomes a large negative number as 
$n_T$ increases.  The problem is compounded because the global 
constraint (\ref{cons2}) requires that we find at least one solution 
with nonpositive magnetic charge.  These considerations
indicate that smaller values of $n_T$, such
as $n_T=2$, where we would have $n_H-n_V=30g-35$,
are more likely to lead to reasonable solutions than
larger values of $n_T$.  But even for the case $n_T=2$, the problem of
systemetizing the factorization criterion analogous to
(\ref{factor8}) becomes comparably unweildy since there are 
many more variables.  We have been unable to find a
solution to the factorization problem for the case $n_T=2$ which also 
satisfies $n_H-n_V=30g-35$.  It is also apparent that seeking solutions
for $n_T\ge 3$ would be a computational morass unlikely
to yield interesting solutions.

\vspace{.1in}
\noindent
{\it Hidden instantons:}\\[.1in]
One aspect which we have not emphasized in our analysis involves the
possible scenarios described in the paragraph following equation
(\ref{e8term}) whereby the $\a$ projection can be realized in
nontrivial ways to break the $E_8$ gauge group to $E_7\times SU(2)$
or $Spin(16)$, thereby describing the effect of ``hidden instantons"
on the fixed points. These possibilities can be analyzed in a manner
very similar to that which we have presented.  The essential difference
is that the contribution $I_8(E_8)_i$ given in (\ref{e8term}) is
replaced with an analogous contribution $I_8(E_7\times SU(2))_i$ or
$I_8(Spin_{16})_i$, which are straightforward to define and to compute.
When we repeat the above analysis for these cases we do not find 
interesting solutions involving nonpositive $g$ for the $Spin(16)$
case.  However, the $E_7\times SU(2)$ case is very intriguing,
but involves a different set of puzzles which, at this time, are
sufficiently muddy that we should avoid further expansion on the subject.
This involves work-in-progress.  

\section{Conclusions}
The precise characterization of anomalies in general 
situations involving {\it M}-theory
orbifolds involves an interesting array of effects.  On the one hand 
are the quantum anomaly and also the various inflow contributions 
enabled by 
modifications to the $dG$ Bianchi identity and the standard 
Chern-Simon's terms in eleven-dimensional supergravity.  These 
two contributions alone are sufficient to understand anomaly 
cancellation in the simplest cases, such as the {\it M}-fivebrane
and orbifolds which only break half of the
bulk supersymmetry.  In situations involving more supersymmetry
breaking, things are more involved.  When there are intersecting
orbifold planes, an ``I-brane" effect occurs which involves an
interplay between electric sources of $G$ localized on one
hyperplane and magnetic sources of $G$ localized on the
intersecting plane.  Finally, local (twisted) tensor fields
are generally needed to supply a local Green-Schwarz mechanism
to cancel against the quantum anomaly, the inflow anomaly and
also the I-brane anomaly.

We have presented, in detail, a particular example corresponding
to an $S^1/{\bf Z}_2\times T^4/{\bf Z}_2$ orbifold which has
seven- and ten-dimensional orbifold planes which interesect at
additional six-dimensional planes.  In this case, it is
shown how local anomaly cancellation on the six-planes requires 
$SO(8)$ gauge matter propagating on each of the
sixteen seven-planes.  But there remains an
unresolved paradox associated with this situation, which is that
it requires $n_T=1/2$. A related
problem had been
previously noted by other authors \cite{font, am} in the context
of the heterotic string.  The
authors of \cite{font} have independently indicated a need for sixteen
{\bf 28}s of $SO(8)$.  The analysis in this paper complements
the results of that paper by offering an alternative 
{\it M}-theoretic and local explanation for these same factors, 
associating them with seven-dimensional submanifolds.

The $S^1/{\bf Z}_2\times T^4/{\bf Z}_2$ orbifold represents
a particular degeneration of $S^1/{\bf Z}_2\times K3$ 
corresponding to the singular ${\bf Z}_2$ orbifold limit of
the $K3$ factor. In \cite{font} the authors also
analyzed a separate case corresponding to the singular
${\bf Z}_3$ orbifold limit of the $K3$ factor.
In that case, they did {\it not} find the
same peculiarities present in the ${\bf Z}_2$ case.  Since that 
${\bf Z}_2\times {\bf Z}_3$ construction also has 
intersecting orbifold planes of precisely the same dimensionality 
as the one featured in this paper, it would be most interesting
to repeat our analysis in that context.  This is also work-in-progress.  

A possibility suggested by this paper is that
that gravitational anomaly cancellation on ten-, six- and two-dimensional
orbifold planes within complicated $T^7$ orbifolds, for example,
involving {\it four}-dimensional fixed planes and/or
intersections would require gauge groups and particle spectra 
which would have relevance to realistic models.  This would 
bring gravitational anomalies into four-dimensional physics
in a novel way.

\pagebreak

\setcounter{equation}{0}
\renewcommand{\thesection}{Appendix A: Tables}
\renewcommand{\theequation}{A.\arabic{equation}}
\section{ }
\renewcommand{\thesection}{A}
In this appendix we compile some results and identities from group theory which are necessary to undertake the detailed analysis described in section 6.

\vspace{.6in}

\begin{center}
\begin{tabular}{|c||cccc|}
\hline
&&&&\\[-2mm]
& Tr & tr & ${\rm tr}_*$ & ${\rm tr}_S$  \\[.1in]
\hline
&&&&\\
SU(n) & $\ft12 n^2-1$ & $n$ & & \\[.2in]
SO(n) & $\ft12 n(n-1)$ & $n$ & & $2^{(N-2)/2}$ \\[.2in]
Sp(n) & $n(2n+1)$ & $2n$ & $n(2n-1)$ & \\[.2in]
$G_2$ & 14  & 7  & & \\[.2in]
$F_4$ & 52  & 26 & & \\[.2in]
$E_6$ & 78  & 27 & & \\[.2in]
$E_7$ & 133 & 56 & & \\[.2in]
$E_8$ & 248 &    & & \\[.1in]
\hline
\end{tabular}\\[4mm]
\parbox{3.5in}{Table 2: Representation dimensions used in the definitions
of the various trace operations.  The trace ${\rm Tr}$ 
refers to the adjoint representation, ${\rm tr}$ refers to the fundamental
representation, ${\rm tr}_*$ refers to the antisymmetric tensor
representation and ${\rm tr}_S$ refers to the spinor representation.}
\end{center}

\pagebreak

\begin{center}
\begin{tabular}{cl}
& \\[1.7in]
SU(n) &
${\rm Tr}\,F^2\,=\,2n\,\,{\rm tr}\,F^2$  \\[.1in]
& ${\rm Tr}\,F^4\,=2n\,\,{\rm tr}\,F^4+6\,({\rm tr}\,F^2\,)^2$ \\[.2in]
& ${\rm tr}\,F^4\,=\ft12\,({\rm tr}\,F^2\,)^2$
\,\,for $SU(2)$ or $SU(3)$ \\[.2in]
SO(n) &
${\rm Tr}\,F^2\,=(n-2)\,{\rm tr}\,F^2$  \\[.1in]
& ${\rm Tr}\,F^4\,=(n-8)\,{\rm tr}\,F^4+3\,({\rm tr}\,F^2)^2$ \\[.1in]
& ${\rm Tr}\,F^6\,=(n-32)\,{\rm tr}\,F^6
  +15\,{\rm tr}\,F^2\,{\rm tr}\,F^4$ \\[.2in]
& ${\rm tr}_S\,F^2\,=2^{(N-8)/2}\,{\rm tr}\,F^2$ \\[.1in]
& ${\rm tr}_S\,F^4\,=-2^{(N-10)/2}\,{\rm tr}\,F^4\,
  +\,3\cdot 2^{(N-14)/2}\,({\rm tr}\,F^2\,)^2$ \\[.2in]
Sp(n) &
${\rm Tr}\,F^2\,=\,(2n+2)\,{\rm tr}\,F^2$ \\[.1in]
& ${\rm Tr}\,F^4\,=\,(2n+8)\,{\rm tr}\,F^4
+3\,({\rm tr}\,F^2)^2$ \\[.2in]
& ${\rm tr}_*\,F^2\,=\,(2n-2)\,{\rm tr}\,F^2$ \\[.1in]
& ${\rm tr}_*\,F^4\,=\,(2n-8)\,{\rm tr}\,F^4
  +3\,({\rm tr}\,F^2)^2$ 
\end{tabular} \\[.3in]
\parbox{4in}{Table 3: Trace relations for the non-exceptional
classical groups. }
\end{center}

\pagebreak

\begin{center}
\begin{tabular}{cl}
&\\[1.2in]
$G_2$ &
${\rm Tr}\,F^2\,=\,4\,{\rm tr}\,F^2$ \\[.1in]
& ${\rm Tr}\,F^4\,=\,\ft52\,({\rm tr}F^2\,)^2$ \\[.2in]
& ${\rm tr}\,F^4\,=\,\frac{1}{4}\,({\rm tr}\,F^2)^2$ \\[.2in]
$F_4$ &
${\rm Tr}\,F^2\,=\,3\,{\rm tr}\,F^2$ \\[.1in]
& ${\rm Tr}\,F^4\,=\,\ft{5}{12}\,({\rm tr}F^2\,)^2$ \\[.2in]
& ${\rm tr}\,F^4\,=\,\ft{1}{12}\,({\rm tr}\,F^2)^2$ \\[.2in]
$E_6$ &
${\rm Tr}\,F^2\,=\,4\,{\rm tr}\,F^2$ \\[.1in]
& ${\rm Tr}\,F^4\,=\,\ft12\,({\rm tr}F^2\,)^2$ \\[.2in]
& ${\rm tr}\,F^4\,=\,\ft{1}{12}\,({\rm tr}\,F^2)^2$ \\[.2in]
$E_7$ &
${\rm Tr}\,F^2\,=\,3\,{\rm tr}F^2$ \\[.1in]
& ${\rm Tr}\,F^4\,=\,\ft16\,({\rm tr}\,F^2)^2$ \\[.2in]
& ${\rm tr}\,F^4\,=\,\ft{1}{24}\,({\rm tr}\,F^2)^2$ \\[.2in]
$E_8$ &
${\rm Tr}\,F^2\,\equiv\,30\,{\rm tr}F^2$ \\[.1in]
& ${\rm Tr}\,F^4\,=\,\frac{1}{100}\,({\rm Tr}\,F^2)^2$ \\[.1in]
& ${\rm Tr}\,F^6\,=\,\ft{1}{7200}\,({\rm Tr}\,F^2\,)^3$ 
\end{tabular}\\[.3in]
\parbox{3in}{Table 4: Trace relations for the exceptional groups.}
\end{center}

\pagebreak

.\vspace{.4in}
\begin{center}
\begin{tabular}{|c||c|c||l|}
\hline
& & & \\[-4mm]
& X & Y & factorization criteria\\
\hline
& & & \\[-2mm]
$SU(N)$ & 0 & $-4/\mu$ & $h=2N/\mu$ \\
& & & \\
$SU(2)$ & $\ft16(h-4/\mu)$ & $\ft13(h-16/\mu)$ & \\
& & & \\
$SU(3)$ & $\ft16(h-6/\mu)$ & $\ft13(h-18/\mu)$ & \\
& & & \\
$SO(N)$ & $q-1/\mu$ & $q-2/\mu$ & $h=(N-8)/\mu+2q$ \\[2mm]
& & & $q\equiv 2^{(N-12)/2}\,h_S$ \\[3mm]
$Sp(N)$ & $h_*+1/\mu$ & $2h_*-2/\mu$ & $h=2N(1/\mu-h_*)+8(1/\mu+h_*)$\\
& & & \\
$G_2$ & $\ft16(h-4/\mu)$ & $\ft16(h-10/\mu)$ & \\
& & & \\
$F_4$ & $\ft16(h-3/\mu)$ & $\ft{1}{18}(h-5/\mu)$ & \\
& & & \\
$E_6$ & $\ft16(h-4/\mu)$ & $\ft{1}{18}(h-6/\mu)$ & \\
& & & \\
$E_7$ & $\ft16(h-3/\mu)$ & $\ft{1}{36}(h-4/\mu)$ & \\
& & & \\
$E_8$ & $-5/\mu$ & $-6/\mu$ & \\
& & & \\
\hline
\end{tabular}\\[4mm]
\parbox{5in}{Table 5: Values of the numbers $X$ and $Y$ ,
which are defined in equation (\ref{xydef}), for the
cases of each individual simple gauge factor, as functions
of the numbers of hypermultiplets transforming
in various representations and as functions of the parameter $\mu$, 
defined in (\ref{mudef}).  The multiplicity $h$ refers to the fundamental
representation whereas $h_*$ and $h_S$ refer to the antisymmetric
tensor representation and spinor representation, respectively.
The right-hand column lists necessary criteria for the factorization
described by the second equation of (\ref{xydef}) as well as the
definition of the parameter $q$.}
\end{center}

\pagebreak

.\vspace{.3in}

\begin{center}
\begin{tabular}{|c||c|}
\hline
& \\[-4mm]
& $n_H-n_V$ \\
\hline
& \\[-2mm]
$SU(N)$ & $(N^2+1)/\mu+s$ \\
& \\ 
$SU(2)$ & $2h-3/\mu+s$ \\
& \\ 
$SU(3)$ & $3h-8/\mu+s$ \\
& \\ 
$SO(N)$ & $\ft12 N(N-15)/\mu+(2N+32)q+s$ \\
& \\
$Sp(N)$ & $N(2N+15)/\mu-h_*\,N(2N-15)+s$ \\
& \\
$G_2$ & $7h+s-14/\mu$ \\
& \\
$F_4$ & $26h+s-52/\mu$ \\
& \\
$E_6$ & $27h+s-78/\mu$ \\
& \\
$E_7$ & $56h+s-133/\mu$ \\
& \\
$E_8$ & $-248/\mu$ \\
& \\
\hline
\end{tabular}\\[4mm]
\parbox{3.5in}{Table 6: Expressions for $n_H-n_V$ for the simple
gauge groups expressed as functions of the representation multiplicities
and as functions of the paramenter $\mu$, defined in (\ref{mudef}).
The multiplicities $h$, $h_*$ and $q$ are defined 
in the caption for
table 5, while $s$ is the number of gauge singlet hypermultiplets.
Note that the factorization criteria 
listed in the right-hand column of table 5 have been used.}
\end{center}

\pagebreak

\setcounter{equation}{0}
\renewcommand{\thesection}{Appendix B: Anomalies in Ten Dimensions}
\renewcommand{\theequation}{B.\arabic{equation}}
\section{ }
\renewcommand{\thesection}{B}
The essential tools used to study orbifold anomalies
are specific polynomials used to describe the anomalies
themselves.  For a given theory, the relevant polynomial can be
readily assembled given the zero-mass spectrum using results from
index theory.  But since each chiral fermion or self-dual tensor
contributes to an anomaly in a linearly independent way, 
it is practical to have ``ready-made" building-block polynomials
associated with each type of field.  In this way, one can readily
determine the anomaly polynomial in a given situation and use this 
to explore the consistency of the theory and as a guide to
additional structure.  In this paper, an emphasis is put on
the nuts-and-bolts aspects of assembling these polynomials 
and using them to ones advantage.   For this reason, and also 
for the reason of making this paper reasonably self-contained,
we include an encyclopedic reveiw of these polynomials in
this and the following appendix.  The exact relationship to 
index theorems is relatively unimportant for our purposes.
So these have been de-emphasized.

The presentation in this appendix paraphrases the more comprehensive
presentation found in \cite{gsw}.  

Ten-dimensional field theories involve three types of
fields which contribute to anomalies at one loop.  These are
chiral spin 3/2 fermions, chiral spin 1/2 fermions, and self-dual 
(or anti self-dual) five-forms.  The total anomaly can be deduced
via descent equations from a formal twelve-form.  Following are
master formulae for deducing the twelve form from a given theory.
The first of these gives the contribution to purely gravitational
anomalies due to a chiral spin 3/2, chiral spin 1/2, 
and self-dual five-form field, respectively,
\brr I_{{\rm GRAV}}^{(3/2)}(R) &=& 
     \frac{1}{(2\pi)^5\,6!}\,\bpl\,
     \frac{55}{56}\,{\rm tr}\,R^6
     -\frac{75}{128}\,{\rm tr}\,R^4\,{\rm tr}R^2
     +\frac{35}{512}\,({\rm tr}\,R^2)^3\,\bpr
     \nonumber\\[.1in]
     I_{{\rm GRAV}}^{(1/2)}(R) &=& 
     \frac{1}{(2\pi)^5\,6!}\,\bpl\,
     -\frac{1}{504}\,{\rm tr}\,R^6
     -\frac{1}{384}\,{\rm tr}\,R^4\,{\rm tr}R^2
     -\frac{5}{4608}\,({\rm tr}R^2)^3,\bpr
     \nonumber\\[.1in]
     I_{{\rm GRAV}}^{({\rm 5-form})}(R) &=&
     \frac{1}{(2\pi)^5\,6!}\,\bpl\,
     -\frac{496}{504}\,{\rm tr}\,R^6
     +\frac{7}{12}\,{\rm tr}\,R^4\,{\rm tr}\,R^2
     -\frac{5}{72}\,({\rm tr}\,R^2)^3\,\bpr \,.
\label{grav10}\err
In these expressions, the Riemann tensor is regarded as an $SO(9,1)$-valued
two-form, $(R_{\mu\nu})_a\,^b$.  The trace is over the $SO(9,1)$
indices $a,b$, and the coordinate indices are 
suppressed.  Wedge products are assumed.  We note that these
formulae are additive for {\it each} field of a given type.  
For instance, relevant theories contain a number of chiral spin 1/2
fields living in vector multiplets.  The 
anomaly due to chiral gauginos would be the second
equation of (\ref{grav10}) times the total number of these gauginos. 

Next are the master formulae for mixed and pure gauge anomalies,
which are due only to chiral spin 1/2 fermions,
\brr I_{{\rm MIXED}}^{(1/2)}(R,F) &=&
     \frac{1}{(2\pi)^5\,6!}\,\bpl\,
     \frac{1}{16}\,{\rm tr}R^4\,{\rm Tr}\,F^2
     +\frac{5}{64}\,({\rm tr}R^2)^2\,{\rm Tr}\,F^2
     -\frac{5}{8}{\rm tr}R^2\,{\rm trace}\,F^4\,\bpr
     \nonumber\\
     I_{{\rm GAUGE}}^{(1/2)}(F) &=&
     \frac{1}{(2\pi)^5\,6!}\,
     {\rm Tr}\,F^6
\label{gauge10}\err
In these expressions, the Yang-Mills field strengths are two-forms
which take values in the adjoint representation
of the gauge group, and ${\rm Tr}$ denotes an adjoint trace.

The above master formulae for $I^{(1/2)}$ and $I^{(3/2)}$ are for chiral
($\Gamma_{11}\psi=\psi$) Weyl spinors.  If the fermions
in question are Majorana-Weyl, which is possible in ten-dimensions,
and have therefore {\it half} of the degrees of freedom of a Weyl spinor,
then the formula  should be multipled by 1/2.  For antichiral spinors
($\Gamma_{11}\psi=-\psi$) the formula should be multiplied by -1.

\subsection{IIB Supergravity}
The ten-dimensional type $IIB$ supergravity theory has a single
self-dual five-form field strength (with analogous four-form potential),
a pair of chiral spin-3/2 Majorana-Weyl gravitinos, and a pair of
antichiral spin-1/2 fermions.  Thus, the total anomaly is described by
\brr I_{12}=I^{(3/2)}_{\rm GRAV}(R)
    -I^{(1/2)}_{\rm GRAV}(R)
    +I^{({\rm 5-form})}_{\rm GRAV}(R) \,.
\label{iib}\err
The coefficients of the $I^{(1/2)}$ and $I^{(3/2)}$ terms each
include a factor of two, since there are two each of the relevant
field types, and also a factor of one-half since the relevant
fields are Majorana-Weyl spinors and therefore have half the
degree of freedom of a Weyl spinor, as described above.  Thus, the overall
coefficients for these terms have absolute value one.
Adding up the various contributions supplied by
equation (\ref{grav10}), we find the  result $I_{12}=0$\,!
Thus, the IIB supergravity theory is anomaly-free.

\subsection{N=1 Supergravity Coupled to Yang-Mills Matter}
The fermionic fields of the D=10 N=1 supergravity multiplet 
comprise a chiral Majorana-Weyl spin-3/2 gravitino and an antichiral 
Majorana-Weyl spin-1/2 dilatino.  There are no (anti) self-dual
5-forms.  This multiplet couples to Yang-Mills supermultiplets
which contains chiral Majorana-Weyl spin-1/2 gauginos living in the
adjoint representation of some gauge group ${\cal G}$.  Thus, the total
anomaly is described  by
\brr I_{12}=\ft12\bpl\,
    I_{{\rm GRAV}}^{(3/2)}(R)
    -I_{{\rm GRAV}}^{(1/2)}(R)\,\bpr
    +\ft12 \bpl n\,I_{{\rm GRAV}}^{(1/2)}(R)
    +I_{{\rm MIXED}}^{(1/2)}(R,F)
    +I_{{\rm GAUGE}}^{(1/2)}(F)\,\bpr \,,
\label{d10anom}\err
where $n={\rm dim}({\cal G})$.
Adding up the various contributions, we then arrive at the total
anomaly polynomial for a generic super Yang-Mills theory
coupled to $D=10$ $N=1$ supergravity,
\brr I_{12} &=& 
     \frac{1}{2\,(2\pi)^5\,6!}\,
     \bpl\,\frac{496-n}{504}\,{\rm tr}\,R^6
     -\frac{224+n}{384}\,{\rm tr}R^4\,{\rm tr}R^2
     +\frac{5}{4608}(64-n)\,({\rm tr}R^2)^3
     \nonumber\\[.1in]
     & & \hspace{.6in}
     +\frac{1}{16}\,{\rm tr}R^4\,{\rm Tr}\,F^2
     +\frac{5}{64}\,({\rm tr}R^2)^2\,{\rm Tr}\,F^2
     -\frac{5}{8}{\rm tr}R^2\,{\rm Tr}\,F^4
     +{\rm Tr}\,F^6\,\bpr
\label{anom10}\err
To cancel this anomaly via a Green-Schwarz mechanism,
it is necessary that the twelve-form factorize into the product
of a four-form and an eight-form.  For a judicious choice of gauge
group, it is possible that ${\rm Tr}F^6$ factorizes into 
a linear combination of ${\rm Tr}F^2\,{\rm Tr}F^4$ and
$({\rm Tr}F^2)^3$.  But $SO(9,1)$ does not enable such a
factorization of ${\rm tr}\,R^6$; this piece must vanish
identically. Therefore $n=496$.  In this case, (\ref{anom10}) becomes
\brr I_{12} &=& 
     \frac{1}{2\,(2\pi)^5\,6!}\,
     \bpl\,-\frac{15}{8}\,{\rm tr}R^4\,{\rm tr}R^2
     -\frac{15}{32}\,({\rm tr}R^2)^3
     +\frac{1}{16}\,{\rm tr}R^4\,{\rm Tr}\,F^2
     \nonumber\\[.1in]
     & & \hspace{1in}
     +\frac{5}{64}\,({\rm tr}R^2)^2\,{\rm Tr}\,F^2
     -\frac{5}{8}{\rm tr}R^2\,{\rm Tr}\,F^4
     +{\rm Tr}\,F^6\,\bpr \,.
\label{form}\err
There is only one possibility to factorize this result into the
product of a two-form and an eight form, which requires
the following property to be satisfied by ${\cal G}$,
\brr {\rm Tr}F^6=
    \ft{1}{48}\,{\rm Tr}F^4\,{\rm Tr}F^2
    -\ft{1}{14400}\,({\rm Tr}F^2)^3 \,.
\label{property}\err
There are exactly two 496-dimensional nonabelian Lie groups with this property,
$SO(32)$ and $E_8\times E_8$.
Given the property (\ref{property}), the anomaly polynomial
(\ref{form}) may be expressed as
\brr I_{12}=-\frac{15}{2\,(2\pi)^5\,6!}\,
    ({\rm tr}R^2-\ft{1}{30}{\rm Tr}F^2)\wedge X_{(8)} \,.
\err
where $X_{(8)}$ is an eight-form given by the following expression,
\brr X_{(8)}=\ft18\,{\rm tr}R^4
    +\ft{1}{32}\,({\rm tr}R^2)^2
    -\ft{1}{240}\,{\rm tr}R^2{\rm Tr}F^2
    +\ft{1}{24}\,{\rm Tr}F^4
    -\ft{1}{7200}({\rm Tr}F^2)^2 \,.
\label{xx10}\err

For $SO(32)$ there is an identity ${\rm Tr}=30\,{\rm tr}$,
and for $E_8$ a similar identity defines the operation
${\rm tr}$. Therefore,, in both cases we can rewrite (\ref{xx10}) as 
\brr X_8=\ft18\,{\rm tr}R^4
    +\ft{1}{32}\,({\rm tr}R^2)^2
    -\ft{1}{8}\,{\rm tr}R^2{\rm tr}F^2
    +\ft{5}{4}\,{\rm tr}F^4
    -\ft{1}{8}({\rm tr}F^2)^2 \,.
\err

\setcounter{equation}{0}
\renewcommand{\thesection}{Appendix C: Anomalies in Six Dimensions}
\renewcommand{\theequation}{C.\arabic{equation}}
\section{ }
\renewcommand{\thesection}{C}
Six-dimensional field theories also involve three types of
fields which contribute to anomalies at one loop.  These are
chiral spin 3/2 fermions, chiral spin 1/2 fermions, and self-dual 
(or anti self-dual) three-forms.  The total anomaly can be deduced
via descent equations from a formal eight-form.  Following are
master formulae for deducing the eight form from a given theory.
The first of these gives the contribution to purely gravitational
anomalies due to a chiral spin 3/2, chiral spin 1/2, 
and a self-dual three-form field, respectively  
\brr I_{{\rm GRAV}}^{(3/2)}(R) &=& 
     \frac{1}{(2\pi)^3\,4!}\,\bpl\,
     -\frac{49}{48}\,{\rm tr}R^4
     +\frac{43}{192}\,({\rm tr}R^2)^2,\bpr
     \nonumber\\[.1in]
     I_{{\rm GRAV}}^{(1/2)}(R) &=& 
     \frac{1}{(2\pi)^3\,4!}\,\bpl\,
     -\frac{1}{240}\,{\rm tr}R^4
     -\frac{1}{192}\,({\rm tr}R^2)^2\,\bpr
     \nonumber\\[.1in]
     I_{{\rm GRAV}}^{({\rm 3-form})}(R) &=&
     \frac{1}{(2\pi)^3\,4!}\,\bpl\,
     -\frac{7}{60}\,{\rm tr}R^4
     +\frac{1}{24}\,({\rm tr}R^2)^2\,\bpr \,.
\label{grav6}\err
In these expressions the Riemann tensor is regarded as an $SO(5,1)$-valued
two-form, $(R_{\mu\nu})_a\,^b$.  The trace is over the $SO(5,1)$
indices $a,b$, and the coordinate indices are 
suppressed.  Wedge products are assumed.  We note that these
formulae are additive for {\it each} field of a given type.  
For instance, relevant theories contain a number of chiral spin 1/2
fields living in vector multiplets.  The contribution
to the total anomaly due to chiral gauginos would be the second
equation of (\ref{grav10}) times the total number of these gauginos.         

Next are the master formulae for mixed and pure gauge anomalies,
which are due only to chiral spin 1/2 fermions,
\brr I_{{\rm MIXED}}^{(1/2)}(R,F) &=&
     \frac{1}{(2\pi)^3\,4!}\,\bpl\,
     \frac{1}{4}\,{\rm tr}R^2\,{\rm trace}\,F^2\,\bpr
     \nonumber\\
     I_{{\rm GAUGE}}^{(1/2)}(F) &=&
     \frac{1}{(2\pi)^3\,4!}\,\bpl
     -{\rm trace}\,F^4\,\bpr
\label{rf6}\err
In these expressions, the Yang-Mills field strengths are two-forms
which take values according to whichever group representation the
gauge fields transform in.       

\subsection{$D=6\,,\,N=2$}
In six dimensions, there are two distinct $N=2$ supergravity multiplets,
one chiral and the other non-chiral
\footnote{By $N=2$ we refer to theories
with {\it twice} the number of supercharges in the minimal 
$D=6$ supergravity theory.  Since the minimal, or $N=1$ theory 
in $D=6$ has four complex supercharges, it follows that
$D=6, N=2$ supergravity theories have eight complex supercharges.
However, since eight complex supercharges
coincides with the $N=4$ theory in four dimensions, these theories
are sometimes (not in this paper) called $N=4$.}.
The chiral supergravity multiplet is denoted $N=2b$ and 
comprises a sechsbein, five self-dual two-forms $B_{IJ}^{(+)}$
(ie: the three-form field strengths satisfy $H=*H$),
and two chiral spin 3/2 gravitinos.  This multiplet can couple only
to $N=2$ tensor multiplets, which each comprise five real scalars, a single
anti self-dual two-form, and two antichiral spin 1/2 gauginos.

For the $N=2b$ supergravity multiplet coupled to $n$ $N=2$ tensor 
multiplets, the quantum anomaly is characterised by the following eight-form,
\brr I_8=\bpl\,2I_{{\rm GRAV}}^{(3/2)}(R)
    +5I_{\rm GRAV}^{({\rm 3-form})}(R)\,\bpr
    -n\bpl\,2I_{GRAV}^{(1/2)}(R)
    +I_{GRAV}^{({\rm 3-form})}(R)\,\bpr \,.
\label{2bt}\err
There are no mixed or pure gauge contributions since there are 
no spin-1 gauge fields in the theory.
The coefficients of the various contributions in (\ref{2bt}) follow from 
the field content of the multiplets specified in the preceeding
paragraph.  The first two terms in (\ref{2bt}) are the contribution from
the $N=2b$ supergravity multiplet while the second two terms
are the contributions from the tensor multiplets.
Using the formulae in (\ref{grav6}) we 
determine that 
\brr I_8(R) &=& \frac{1}{(2\pi)^3\,4!}
    \frac{n-21}{8}\,\bpl\,
    {\rm tr}R^4-\ft14\,({\rm tr}R^2)^2 \,\bpr\,.
\err
Since ${\rm tr}R^4$ cannot factorize, the first term in this
expression must vanish if the theory is to be anomaly-free.
This then requires that $n=21$\,. 

\subsection{$D=6\,,\,N=1$}
In six dimensions there is only one supergravity multiplet
with $N=1$ supersymmetry.  This multiplet is chiral and comprises
a sechsbein, a single self-dual two-form and a
chiral spin-3/2 gravitino.  There are three distinct matter multiplets
to which this multiplet can couple.  These are the vector multiplet
which includes a spin-1 gauge field and a chiral spin-1/2 gaugino,
the hypermultiplet which includes four real scalars and an antichiral 
spin-1/2 fermion, and the tensor multiplet which includes a single 
real scalar, a single anti self-dual two-form and an  antichiral 
spin-1/2 fermion.
Each of these multiplets contributes to a gravitational anomaly. 
To evaluate a potential gauge anomaly, we have to specify the 
group representation relevent to each multiplet.  We restrict to the
case where vector multiplets transform in the adjoint and tensor
multiplets are gauge singlets.  The representation of hypermultiplets
can be chosen freely.

To begin, we restrict to the case where $G$ is simple.  We will
generalize this to the case where $G$ is semi-simple below.
Thus, given a gauge group, the only freedom we allow is in the choice of 
representation for the hyper multiplets, some of which can be 
gauge singlets, and the number of gauge-singlet tensor multiplets. 
For the case of perturbative string effective theories, $n_T=1$.  
Including nonperturbative effects can change this, however.  
Similarly, {\it M}-Theory also gives rise to $n_T\ne 1$ effective
theories.

If we include $n_V$ vector multiplets, $n_H=\sum_\a n_\a$ hyper multiplets
with $n_\a$ hypermultiplets in the representation $R_\a$,
and $n_T$ tensor multiplets, then the total anomaly is described by the
following eight-form,
\brr I_8 &=& I^{(3/2)}_{\rm GRAV}(R)
    +(\,n_V-n_H-n_T\,)\,I^{(1/2)}_{\rm GRAV}(R)
    +(\,1-\,n_T\,)\,I^{({\rm 3-form})}_{\rm GRAV}(R)
    \nonumber\\[.1in]
    & & +\bpl\,I_{\rm MIXED}^{(1/2)}(R,F)
    +I_{\rm GAUGE}^{(1/2)}(F)\,\bpr_{\rm ADJ}
    \nonumber\\[.1in]
    & & -\sum_\a n_\a\,\bpl\,I_{\rm MIXED}^{(1/2)}(R,F)
    +I_{\rm GAUGE}^{(1/2)}(F)\,\bpr_{R_\a}
\err
which follows directly from the discussion above, given the field content
of the various multiplets.  Note that the subscripts ADJ and $R_\a$
refer to the representations being traced over in the respective
anomaly polynomial, and that $n_\a$ is the number of hypermultiplets
in the representation $R_\a$\,.
Using the formulae in (\ref{grav6}),
we then compute
\brr I_8 &=& \frac{1}{(2\pi)^3\,4!}\,\bpl\,
    \hspace{1mm}\frac{1}{240}\,(\,n_H-n_V+29 n_T-273\,)\,{\rm tr}R^4 
    \nonumber\\
    & & \hspace{15mm}
    +\frac{1}{192}(\,n_H-n_V-7 n_T+51\,)\,(\,{\rm tr}\,R^2\,)^2
    \nonumber\\[.1in]
    & & \hspace{15mm}
    +\frac{1}{4}\,{\rm tr}R^2\wedge({\rm Tr}\,F^2
    -\,\sum_\a\,n_\a\,{\rm tr}\,F_\a^2\,)
    \nonumber\\[.1in]
    & & \hspace{15mm}
    -(\,{\rm Tr}\,F^4
    -\,\sum_\a\,n_\a\,{\rm tr}\,F_\a^4\,)\,\bpr\,.
\label{a6n1}\err
A more precise description of the traces over the gauge group
representations is given below.
We require that the anomaly (\ref{a6n1})
{\it factorize} so that the anomaly
can be canceled locally by a Green-Schwarz mechanism.  
This requires that the coefficient of the first term in
(\ref{a6n1}) vanishes, as this term is irreducible
(ie: it is impossible to factorize ${\rm Tr}R^4$).  We thus
determine the following requirement
\brr n_H-n_V+29\,n_T=273\,.
\label{6dr}\err
For the case of perturbative heterotic string compactifications, 
one finds generically that $n_T=1$, 
since there is only one tensor in the relevant effective theory.  
In that case, equation (\ref{6dr}) 
reduces to $n_H-n_V=244$, a commonly cited string requirement.
Note that in {\it M}-theory we expect more than a single tensor
field since the eleven-dimensional three-form $C$ can provide us
with several two-forms upon dimensional reduction.  In addition,
fivebranes, which are important ingredients in {\it M}-theory,
provide additional two-forms since their dynamics involve
six-dimensional tensor multiplets.

We impose (\ref{6dr}).  Thus we can reexpress the anomaly 
(\ref{a6n1}) as follows,
\brr I_8 &=& \frac{1}{(2\pi)^3\,4!}\,\frac{3}{2}\,\bpl\,
    \frac{9-n_T}{8}\,(\,{\rm tr}\,R^2\,)^2
    +\frac{1}{6}\,{\rm tr}R^2\wedge({\rm Tr}\,F^2
    -\,\sum_\a\,n_\a\,{\rm tr}\,F_\a^2\,)
    \nonumber\\[.1in]
    & & \hspace{1.9in}
    -\frac{2}{3}(\,{\rm Tr}\,F^4
    -\,\sum_\a\,n_\a\,{\rm tr}\,F_\a^4\,)\,\bpr\,.
\label{prov}\err
This expression needs some care to be evaluated properly, 
especially if semisimple groups are allowed. 
If the gauge group involves $N$ simple factors,
$G_1\times G_2\times\cdots G_N$ and if 
the hypermultiplets trasform
as $(R_1,R_2,...,R_N)$ then it turns out that
${\rm tr}F^2=\sum_\a n_\a {\rm tr}\,F_{\a}^2$ and
${\rm tr}F^4=\sum_\a{\rm tr}F_\a^4
+6\sum_{\a<\b}\,n_{\a\b}{\rm tr}F_\a^2\wedge{\rm tr}F_\b^2$ where
$n_\a$ is the number of multiplets transforming
as $R_\a$, and $n_{\a\b}$ are the number of multiplets transforming as 
$(R_\a,R_\b)$ under the $G_\a\times G_\b$ subgroup.  
For example, in the case of two gauge factors 
$G_1\times G_2$, we would find 
$n_1={\rm dim}R_2$, $n_2={\rm dim}R_1$ and $n_{12}=1$.  
For vector multiplets transforming in the adjoint, we have
the relation ${\rm Tr}\,F^n=\sum_\a\,{\rm Tr}\,F_\a^n$.

Using the relationships discussed above, we generalize (\ref{prov}) 
to the case of semisimple gauge group $G_1\times G_2\times\cdots G_N$
with the representation structure described above, and find the
following anomaly polynomial,
\brr I_8 = \frac{1}{(2\pi)^3\,4!}\,\frac{3}{2}\,\bpl\,\,
    \frac{9-n_T}{8}\,(\,{\rm tr}\,R^2\,)^2
    +\frac{1}{6}\,{\rm tr}\,R^2\wedge\sum_\a\,X_\a^{(2)}
    -\frac{2}{3}\,\sum_\a\,X_\a^{(4)}
    +4\,\sum_{\a<\b}\,Y_{\a\b} \,\bpr \,,
\label{aa6}\err
where the following abbreviations have been used,
\brr X_\a^{(2)} &=& {\rm Tr}\,F_\a^2-n_\a\,{\rm tr}\,F_\a^2
     \nonumber\\[.1in]
     X_\a^{(4)} &=& {\rm Tr}\,F_\a^4-n_\a\,{\rm tr}\,F_\a^4
     \nonumber\\[.1in]
     Y_{\a\b} &=& n_{\a\b}\,{\rm tr}\,F_\a^2\wedge {\rm tr}\,F_\b^2 \,.
\label{nnn}\err
The form of the anomaly polynomial (\ref{aa6}) was presented
in \cite{schwarz6}, with the same conventions used here,
but for the special case $n_T=1$.  Note that the terms 
$n_\a\,{\rm tr}\,F_\a^2$, $n_\a\,{\rm tr}\,F_\a^4$ have an implicit
sum over the different representations which might be included,
and that $n_\a$ therefore represents a set of multiplicities, one
for each such representation.  A similar statement applies
to the definition of $Y_{\a\b}$.

\renewcommand{\thesection}{Acknowledgements}
\section{ }
M.F. would like to thank Jens Erler and Andr{\'e} Mimiec
for helpful discussions.

\end{document}